\documentclass[useAMS,usenatbib]{mn2e}
\usepackage{graphicx}                   
\usepackage{float}
\usepackage{amsmath}
\usepackage{amssymb}
\usepackage{multirow}
\usepackage{upgreek}
\usepackage{bm}
\usepackage{color}
\input{epsf.sty}                        
\input{psfig.sty}
\date{}

\begin{document}

\title[FRB-like burst and glitch]
{Constraining mechanism associated with fast radio burst and glitch from SGR J1935}

\author[Wang et al.]{Wei-Hua Wang$^{1,2}$\thanks{E-mail: wang-wh@pku.edu.cn}, Heng Xu$^{3}$, Wei-Yang Wang$^{1}$, Shuang Du$^{1}$, Quan Cheng$^{4}$,
\and
Xiao-Ping Zheng$^{4,5}$ and Ren-Xin Xu$^{1,3}$\thanks{E-mail:r.x.xu@pku.edu.cn}\\
$^1$School of Physics, Peking University, Beijing 100871, China\\
$^2$Key Laboratory of Quark and Lepton Physics(MOE), Central China Normal University, Wuhan 430079, China\\
$^3$Kavli Institute for Astronomy and Astrophysics at Peking University, Beijing 100871, China\\
$^4$Institute of Astrophysics, Central China Normal University, Wuhan 430079, China\\
$^5$School of Physics, Huazhong University of Science and Technology, Wuhan 430079, China\\
}

\maketitle

\begin{abstract}
The discovery of fast radio burst (FRB) 200428 from galactic SGR J1935+2154 makes it possible to measure rotational changes accompanied by FRBs and to test several FRB models which may be simultaneously associated with glitches.
Inspired by this idea, we present order of magnitude calculations to the scenarios proposed.
FRB models such as global starquakes, crust fractures and collisions between pulsars and asteroids/comets are discussed.
For each mechanism, the maximum glitch sizes are constrained by the isotropic energy release during the X-ray burst and/or the SGR J1935+2154-like radio burst rate.
Brief calculations show that, the maximum glitch sizes for different mechanisms differ by order(s) of magnitude.
If glitches are detected to be coincident with FRBs from galactic magnetars in the future, glitch behaviors (such as glitch size, rise timescale, the recovery coefficient and spin down rate offset) are promising to serve as criterions to distinguish glitch mechanisms and in turn to constrain FRB models.

\end{abstract}


\section{Introduction}
\label{introduction}
Fast radio bursts (FRBs) are millisecond-timescale pulses observed in radio band originated from sources at extra-galactic distances~\citep{2007Sci...318..777L,2013Sci...341...53T,2016Natur.531..202S}, except the recently reported FRB 200428 from the direction of the galactic soft gamma-ray repeater (SGR) J1935+2154~\citep{2020Natur.587...54C,2020Natur.587...59B,2020Natur.587...63L}.
The known population of FRBs comes from $118$ independent sources~\footnote{http://frbcat.org}, and can be divided into one-off bursts~\citep{2007Sci...318..777L,2013Sci...341...53T} and repeating bursts~\citep{2016Natur.531..202S,2020Natur.582..351C,2020ApJ...896L..41C}, it is still unclear whether all FRBs will eventually repeat.
No other high-energy counterpart associated with FRBs has been detected except the peculiar double-peaked X-ray burst coincident with FRB 200428.
The physics behind these mysterious violent radio bursts is still under heated debate, dozens of models regarding different generators have been proposed~(see \cite{2018PrPNP.103....1K},~\cite{2019A&ARv..27....4P} and~\cite{2020Natur.587...45Z} for a recent review, and references therein), most of which involve neutron stars (NSs) or even magnetars considering the timescale and energy requirements.
The recent discovery of spatial and temporal association between FRB 200428 and galactic SGR J1935+2154 strongly support its magnetar origin for at least part of the FRBs~\citep{2020Natur.587...54C}.
The analysis of magnetoionic environments of FRBs even support magnetars behind all FRBs (repeating and non-repeating) as a plausible possibility, except for the source of FRB 121102 which requires extreme conditions under the magnetar model~\citep{2020MNRAS.499..355W}.

The galactic origin and association with the peculiar X-ray burst makes FRB 200428 from SGR J1935+2154 a promising target to study the physical mechanism behind.
Emission properties of the peculiar non-thermal X-ray counterpart of galactic FRB 200428 are presented here.
On 2020 April 28, a short, doubled-peaked radio burst, FRB 200428, was reported by~CHIME/FRB and STARE independently~\citep{2020Natur.587...54C,2020Natur.587...59B}.
Simultaneously, a non-thermal X-ray burst was observed with INTEGRAL~\citep{2020ApJ...898L..29M}, AGILE~\citep{2021NatAs.tmp...31T}, Konus-\textit{Wind} (KW)~\citep{2021NatAs.tmp...30R} and \textit{Insight}-HXMT~\citep{2021NatAs.tmp...48L}.
According to Ridnaia et al., the burst total energy fluence (in the $20-250~\rm keV$) during a total duration of $0.464~\rm s$ is $(9.7\pm 1.1)\times 10^{-7}~\rm erg/cm^{2}$~\citep{2021NatAs.tmp...30R}, therefore, the isotropic burst energy release in X-rays is $E_{\rm iso-X}\simeq 1.148\times 10^{40}D_{10}^{2}~\rm erg$.
$D_{10}=D/(10~\rm{kpc})$, $D$ is the distance to SGR J1935+2154.
The SGR J1935+2154-like radio burst rate is estimated to be $(0.007-0.04)~\rm{yr^{-1}}$ magnetar$^{-1}$~\citep{2020Natur.587...54C}.

The galactic origin of FRB 200428 has inspired us that, it is possible to detect the temporal association between FRB and magnetar's rotational change (glitch or anti-glitch) in the future through multi-frequency observations (X-ray and radio).
Different glitch triggering mechanisms result in glitches differ by orders of magnitude.
If a glitch is confirmed to be in association with an FRB, the glitch size may in turn serve as a constraint on mechanisms of FRBs.
Therefore, in this paper, we focus on the subclass of glitch-related FRB models, discussing the glitch size reachable with the constraints from the X-ray burst associated with FRB 200428 and/or the burst rate.

Previously, possible association between glitch and FRBs has been discussed.
Wang et al. studied the energy and waiting time distributions of repeater FRB 121102, the similarities to earthquake manifestations may suggest their origins from pulsar starquakes~\citep{2018ApJ...852..140W}.
Dado \& Dar argued that glitch-associated mini contractions within slow rotators may result in highly relativistic dipolar $e^{+}e^{-}$ bunches lunched from the polar caps which then emit FRBs~\citep{2020arXiv200708370D}.
Besides, Suvorov \& Kokkotas found that, magnetic deformation may dominate over spin down strain in the crust of magnetars, reconfiguration of the strong and highly multipolar crustal magnetic fields generates zones of magnetic stress, which is potentially strong enough to facilitate frequent crust cracks and result in local crustquakes~\citep{2019MNRAS.488.5887S}.
Apart from starquake, interactions between pulsars and small celestial bodies such as asteroids/comets or asteroid belts may also result in FRBs and glitches (or anti-glitches) simultaneously~\citep{2015ApJ...809...24G,2016ApJ...829...27D,2020ApJ...897L..40D}.

Calculations hereafter are based on rotational parameters of SGR J1935+2154.
It has measured spin period and period derivative of $P=3.24~\rm{s}$ and $\dot{P}=1.43(1)\times10^{-11}~\rm{s~s^{-1}}$, a spin-inferred dipolar magnetic field strength of $B_{d}\sim 2.2\times 10^{14}~\rm{G}$ and a spin down luminosity of $L_{\rm sd}\sim 1.7\times 10^{34}~\rm{erg/s}$~\citep{2016MNRAS.457.3448I}.
No persistent radio emission was reported at present~\citep{Fong2014}.
Uncertainties in the last quoted digit are given in parentheses.
G\"o\u{g}\"u\c{s} et al. argued a local surface magnetic field strength of $9.6\times 10^{14}~\rm{G}$ through X-ray spectral modelling of the persistent emission~\citep{2020ApJ...905L..31G}.
The characteristic age of SGR J1935+2154 is $\tau_{\rm c}\sim 3.6~\rm{kyr}$.
However, sky location has placed it coincident with supernova remnant (SNR) G57.2+0.8~\citep{Galensler2014}, whose estimated distance ranges from $1.5-12.5~\rm{kpc}$~\citep{2018ApJ...852...54K,2020ApJ...905...99Z,2020ApJ...898L..29M,2021MNRAS.tmp..771B}.
Recently, kinetic age of SNR G57.2+0.8 hosting SGR J1935+2154 is estimated to be $\tau_{\rm k}\geq 2\times 10^{4}~\rm{yr}$~\citep{2020ApJ...905...99Z}.

Surface temperature ($T_{\rm s}$) of SGR J1935+2154 is crucial, because it reflects the crustal temperature to some extent, which determines whether superfluidity could exist in the crust of this magnetar.
Its spectrum contains a blackbody component with temperature of about $500~\rm{eV}$ plus a power-law component~\citep{2016MNRAS.457.3448I}.
SGR J1935+2154 entered its active episode on 2020 April 27, temperature of the thermal component of the persistent X-ray emission decreases from $1.6~\rm keV$ at the onset of the outburst to $\sim 0.5~\rm keV$ in a couple of days.
However, it is unknown whether the temperature of $\sim 0.5~\rm keV$ represents the global temperature of the star or that of a hot spot on the surface~\citep{2020ApJ...904L..21Y,2021arXiv210401925Y}.
The possible existence of neutron superfluidity in the crust of SGR J1935+2154 are further discussed in section 4.

In this article, based on the hypothesis that it is possible to measure the rotational changes associated with FRBs from galactic magnetars, we explore the glitch size achievable for various mechanisms that possibly result in glitches and FRBs simultaneously.
Since FRB 200428 from SGR J1935+2154 is the only galactic FRB till now, it is also the only FRB found to be associated with high-energy counterpart, we calculate the glitch size based on rotational parameters of SGR J1935+2154, energy release and burst rate of FRB 200428 throughout this paper.
Note that this proposal is general and can be applied to any other FRBs from galactic magnetars in the future.
Rotational parameters of SGR J1935+2154 and some emission properties of FRB 200428 are summarized in Table \ref{SGR J1935+2154 tabel}.
The specific glitch mechanisms include spin down and magnetic strained starquake of NSs, spin down strained starquake of solid quark stars (also named strangeon stars)~\citep{2003ApJ...596L..59X}, interactions between pulsars and small celestial bodies such as asteroids/comets (or asteroid belts) and subsequent vortex unpinning.

This article is organized as follows.
The glitch sizes contributed by pure starquakes (global and local starquakes) are calculated in section 2, the glitch sizes contributed by angular momentum transfer during collisions are calculated in section 3, glitch sizes amplified by vortex unpinning are estimated in section 4.
Conclusions and discussions are present in section 5.

\begin{table}
\begin{center}

    \caption{Parameters of SGR J1935+2154 and FRB 200428}
    \begin{tabular}{ | p{1.7cm} | p{3.0cm} |p{2.0cm}|}
    \hline
    Parameter          & Value                                & Refs.$^a$\\ \hline
    $\nu$              & $0.308~\rm{Hz}$                      & (1) \\
    $\dot\nu$          & $-1.356\times 10^{-12}~\rm{Hz/s}$    & (1) \\
    $B_{d}$            & $2.2\times 10^{14}~\rm{G}$           & (1) \\
    $L_{\rm{sd}}$      & $1.7\times 10^{34}~\rm{erg/s}$       & (1) \\
    $\tau_{\rm c}$     & $3.6~\rm{kyr}$                       & (1) \\
    $\tau_{\rm k}$     & $>20~\rm{kyr}$                       & (6)\\
    $D$ (Distance)     & $(1.5-12.5)~\rm{kpc}$                & (2),(3),(4),(5)  \\
    Burst rate         & $(0.007-0.04)~\rm{yr^{-1}}$ magnetar$^{-1}$ & (7)\\
    $T_{\rm s}$        & ?                                    & (1),(8)\\
    Estimated $E_{\rm{iso-X}}$   & $\sim10^{40}~\rm{erg}$               & (9)\\
    \hline
    \end{tabular}
    \label{SGR J1935+2154 tabel}
\end{center}
$^a$ \scriptsize{References: (1)~\cite{2016MNRAS.457.3448I};(2)~\cite{2018ApJ...852...54K};(3)~\cite{2020ApJ...905...99Z};(4)~\cite{2020ApJ...898L..29M};(5)~\cite{2021MNRAS.tmp..771B};(6)~\cite{2020ApJ...905...99Z};(7)~\cite{2020Natur.587...54C};(8)~\cite{2021arXiv210401925Y};(9)~\cite{2021NatAs.tmp...30R}.} \\
\end{table}

\section{The starquake scenario}
The brief picture of starquakes is presented below.
NSs are born to be hot and own an initial oblateness before solidification.
After their solidification due to cooling, gravity and the loss of centrifugal force during spin down tend to make the star spherical, but the elasticity of the solid crust resists this change and tends to make them oblate.
In this case, NSs with solid crusts are always more oblate than the rotating incompressible fluid stars with the same mass $M$ and radius $R$.
Stress develops during the resistance before it exceeds the critical stress that the crust can support.
Once the stress is relaxed suddenly, a glitch occurs and elastic energy will be released.
Apart from the deformation supported by rotation, effect of magnetic stress is more important in magnetar.
The magnetically strained local starquakes in magnetars will be further discussed in subsections 2.3 and 4.2.

Baym \& Pines first formulated the starquake problem using the model of a self-gravitating, completely solid, incompressible elastic sphere of uniform density and shear modulus within Newtonian gravity framework~\citep{1971AnPhy..66..816B}.
According to this work, the equilibrium oblateness $\varepsilon$ is determined by the competition between gravity, centrifugal force and the elasticity.
The total energy of a NS with solid crusts is
\begin{equation}
\label{eq1}
E_{\rm{total}}=E_{0}+\frac{L^{2}}{2I}+A\varepsilon^{2}+B(\varepsilon-\varepsilon_{0})^{2},
\end{equation}
where $E_{0}$ is the total energy of a non-rotating star, $L$ is the total angular momentum, $I$ is the total moment of inertia, $A\varepsilon^{2}$ is modification of gravitational energy of an ellipsoid relative to a spheroid star with the same mass and density, parameter $B$ measures the elastic strain contribution to the equilibrium configuration of a rotating NS with solid crust, $B=\mu V/2$, $\mu$ is the shear modulus.
For NSs, $V$ represents the volume of the crust, while for strangeon stars, $V=4\uppi R^{3}/3$.
$\varepsilon_{0}$ is the reference oblateness, $B(\varepsilon-\varepsilon_{0})^{2}$ represents the elastic energy.
For a pulsar with mass $M$, radius $R$ and oblateness $\varepsilon$, its moment of inertia is
$I=I_{0}(1+\varepsilon)$, where $I_{0}=(2/5)MR^{2}$ is the moment of inertia of the star without rotation.
For an incompressible solid star,
\begin{equation}
\label{eq3}
A=\frac{3}{25}\frac{GM^{2}}{R}=6.21\times10^{52}M_{1.4}^{2}R_{10}^{-1}~\rm erg,
\end{equation}
where $G$ is the gravitational constant, $M_{1.4}=M/(1.4~M_{\odot})$, $R_{10}=R/(10~\rm{km})$, $M_\odot=1.99\times10^{33}~\rm g$, $1~\rm erg=1~\rm g~cm^{2}~s^{-2}$.
The equilibrium oblateness is obtained by minimizing the total energy in Eq. (\ref{eq1}) with respect to $\varepsilon$, thus
\begin{equation}
\label{eq5}
\varepsilon=\frac{I_{0}\uppi^{2}\nu^{2}}{A+B}+\frac{B}{A+B}\varepsilon_{0},
\end{equation}
$\nu$ is the spin frequency ($\dot\nu$ is the spin down rate).
According to~\cite{1971AnPhy..66..816B}, the reference oblateness $\varepsilon_{0}$ corresponds to the rotation rate when the crust solidifies (about one year after the NS's birth),
\begin{equation}
\label{eq6}
\varepsilon_{0}=I_{0}\uppi^{2}\nu_{0}^{2}/A,
\end{equation}
where $\nu_{0}$ is the spin frequency at the moment crust solidifies.
During the normal spin down state, NSs change their oblateness at the rate
\begin{equation}
\label{eq7}
\dot\varepsilon=\frac{2\uppi^{2}I_{0}\nu\dot\nu}{A+B},
\end{equation}
as a comparison, the reference oblateness of a fluid star changes the oblateness at the rate
\begin{equation}
\label{eq8}
\dot\varepsilon_{0}=\frac{2\uppi^{2}I_{0}\nu\dot\nu}{A}.
\end{equation}
The reference oblateness decreases every time a crust quake occurs.
According to Eq.(\ref{eq5}), the glitch size can be expressed as
\begin{equation}
\label{eq9}
\frac{\Delta\nu}{\nu}=-\frac{\Delta I}{I}=\Delta\varepsilon=\frac{B}{A+B}\Delta\varepsilon_{0}=b\Delta\varepsilon_{0},
\end{equation}
where $\Delta\nu$ is the frequency increase during glitch, $\Delta I (<0) $ is change in the moment of inertia, $\Delta \varepsilon (>0)$ and $\Delta\varepsilon_{0} (>0)$ are changes in oblateness of NSs and the reference fluid star, $b=B/(A+B)$ is the rigidity parameter which was previously estimated to be $b\sim 10^{-5}$ for typical NS parameters~\citep{1971AnPhy..66..816B}.
A large $b$ is conductive to produce large glitches.

Crustquake model of NSs have been developed recently~\citep{2019PASA...36...36G,2020MNRAS.491.1064G,2021arXiv210612604R}, which got challenged compared with observations.
Previously, Baym \& Pines found that, crustquakes of NSs encountered difficulty in explaining the frequent and large glitches ($\Delta\nu/\nu\sim 10^{-6}$) in the Vela pulsar, unless the Vela pulsar is a lightest NS and glitch of size $\sim 10^{-6}$ is an extremely unusual event~\citep{1971AnPhy..66..816B}.
Besides, recent calculations show that, the rigidity parameter $b$ is much smaller than the estimate in~\cite{1971AnPhy..66..816B}.
For example, Cutler et al. found that $b\sim 2\times 10^{-7}$ for a more realistic NS structure with a thin solid crust afloat on an incompressible liquid core~\citep{2003ApJ...588..975C}, which is a factor of $\sim 40$ smaller than that found by Baym \& Pines using a self-gravitating, completely solid, incompressible sphere of constant density and shear modulus~\citep{1971AnPhy..66..816B}.
Zdunik et al. further studied the dependence of parameters of $A$ and $B$ (and thus $b$) on stellar mass and equation of state (EOS).
$A$ was found to be severely underestimated, while $B$ was found to be around $\sim 1.4\times 10^{46}~\rm{erg}$ by employing DH EOS for a NS with mass $M=1.4~M_{\odot}$~\citep{2008A&A...491..489Z}, much smaller than the value $B\sim 10^{47}~\rm{erg}$ employed by Baym \& Pines~\citep{1971AnPhy..66..816B}.
It should be noted that, according to Fig.(15) in their work, for the DH EOS, varying the stellar mass from $1.0~M_{\odot}$ to $2.0~M_{\odot}$ brings decrease in parameter $B$ by only a factor of about $1$, so $B$ is not as sensitive to stellar mass as $A$.
Moreover, Giliberti et al. concluded that, the strain developed between two successive large glitches in the Vela pulsar is impossible to trigger a starquake, unless the sequence of starquakes in a neutron star is an history-dependent process~\citep{2020MNRAS.491.1064G}.
Recently, Rencoert et al. showed that, even the subclass of small glitches cannot be accounted for by the starquake model of neutron stars from the view of glitch activity~\citep{2021arXiv210612604R}.
Apart from all these, in the vortex model~\citep{1977ApJ...213..527A}, whether the crustal moment of inertia of a NS is enough or not is still under debate~\citep{2010ApJ...719L.111Y} especially when the non-dissipative entrainment effect is taken into account~\citep{2012PhRvL.109x1103A,2013PhRvL.110a1101C,2016ApJS..223...16L,2016PhRvL.117w2701W,2017PhRvL.119f2701W}.

An alternative glitch model is starquake of solid quark stars (or strangeon stars)~\citep{2003ApJ...596L..59X,2004APh....22...73Z,2014MNRAS.443.2705Z}.
Two types of glitches are proposed: bulk-invariable (Type I) and bulk-variable ones (Type II)~\citep{2008MNRAS.384.1034P}.
Type I glitches are similar to crust quake of NSs, they occur because of stress accumulation during long-term spin down process.
What makes type I glitches different from crust quakes of NSs is that, the shear modulus of strangeon stars is higher than that of NSs, according to Eq.(\ref{eq9}), the corresponding glitch sizes of type I glitches in strangeon stars will be larger if the time interval between two successive glitches is identical.
\cite{2018MNRAS.476.3303L} further considered the contribution of density uniformity on the moment of inertia, the total moment of inertia of a NS takes the form
\begin{equation}
\label{eq9a}
I=I_{0}(1+\varepsilon)(1+\eta),
\end{equation}
where $\eta (>0)$ measures the level of density uniformity, $\eta$ tends to be smaller and smaller.
The corresponding glitch size is given by
\begin{equation}
\label{eq10}
\frac{\Delta\nu}{\nu}=-\frac{\Delta I}{I}=\Delta\varepsilon+\Delta\eta.
\end{equation}
According to Eq.(25) in \cite{2021MNRAS.500.5336W}, the glitch size and the recovery coefficient $Q$ of a glitch in the strangeon star model have the relation
\begin{equation}
\label{eq11}
\frac{\Delta\nu}{\nu}=\frac{(k-1)\Delta\varepsilon}{Q},
\end{equation}
where $k$ is a constant, $k\sim 5-10$~\citep{2021MNRAS.500.5336W}.
Type II glitches are supposed to be induced by accretion or loss of centrifugal force, the released energy comes from gravitational energy, which represents an energy release huge enough to explain the outbursts of magnetars~\citep{2014MNRAS.443.2705Z}.
For more details about type I and II glitches in the strangeon model, please turn to \cite{2018MNRAS.476.3303L} and \cite{2021MNRAS.500.5336W}.

In this section, we explore the glitch size achievable for starquakes of NSs and strangeon stars (SSs).
Note that, though deformation supported by rotation is negligible compared with magnetic deformation in magnetars, the spin down strained starquakes will still be considered and calculated, for comparison with starquakes in strangeon stars.
In the following, NS represents a neutron-rich pulsar unless specifically illustrated.

\subsection{Spin down strained starquakes in NSs}
For SGR J1935+2154, $\nu=1/P=0.308~\rm{Hz}$, $\dot\nu=-1.356\times10^{-12}~\rm{Hz/s}$.
According to Figs.(4) and (5) in~\cite{2008A&A...491..489Z}, parameter $A$ depends sensitively on NS mass and EOS.
For a NS with typical mass $M=1.4~M_{\odot}$, $A\simeq (2-10)\times 10^{52}~\rm{erg}$.
Parameter $B$ does not sensitively depend on the NS mass, according to Fig.(10) in the same work, $B\simeq 1.4\times 10^{46}~\rm{erg}$ for the DH EOS.
According to Eq.(\ref{eq6}), assuming $\nu_{0}\sim100~\rm{Hz}$, reference oblateness of this star will be
\begin{equation}
\label{eq11ex}
\varepsilon_{0}\simeq 1.1\times 10^{-3}M_{1.4}R_{10}^{2}\nu_{100}^{2}A_{53}^{-1},
\end{equation}
where $\nu_{100}=\nu_{0}/(100~\rm{Hz})$ and $A_{53}=A/(10^{53}~\rm{erg})=0.2-1$.
The oblateness change rate of the pulsar is
\begin{eqnarray}
\label{eq12}
\dot\varepsilon &\simeq &\frac{2\uppi^{2}I_{0}\nu\dot\nu}{A+B}
\simeq -2.9\times10^{-12}M_{1.4}R_{10}^{2}A_{53}^{-1}
\nonumber\\
&&\times (\frac{\nu}{0.308~\rm{Hz}})(\frac{\dot\nu}{-1.356\times 10^{-12}~\rm{Hz/s}})~\rm{yr^{-1}}.
\end{eqnarray}
Shear modulus of a NS is $\mu\sim 10^{30}~\rm{erg/cm^{3}}$~\citep{1969Natur.223..597R,1971AnPhy..66..816B,2001MNRAS.324..811J}.
We set the time interval since last glitch in this magnetar as $\Delta t_{q}$ and estimate $\Delta t_{q}$ from two aspects.
Firstly, Espinoza et al. has calculated the relation between the mean glitching rate ($\dot{N_{\rm g}}$) and pulsar spin down rate using a data base containing $315$ glitches in $102$ pulsars~\citep{2011MNRAS.414.1679E}, the relation is
\begin{equation}
\label{eq13}
\dot{N_{\rm g}}\simeq 3\times 10^{-3}\times (\frac{\dot{\nu}}{10^{-15}~\rm{Hz/s}})^{0.47}~\rm{yr^{-1}}.
\end{equation}
For SGR J1935+2154, its average glitching rate is
\begin{equation}
\label{eq14}
\dot{N_{\rm g}}\simeq 8.9\times 10^{-2}~\rm{yr^{-1}},
\end{equation}
which means an average waiting time of $\Delta t_{q}=1/\dot{N_{\rm g}}\simeq 11~\rm{yr}$.
Secondly, the quake happens every time the stress equals the critical stress value $\sigma_{\rm c}$.
According to Eq.(18) in~\cite{1971AnPhy..66..816B}, from the view of stress accumulation, the  waiting time will be
\begin{equation}
\label{eq15}
\Delta t_{q}=\frac{|\Delta\sigma|}{\dot\sigma},
\end{equation}
where $\dot{\sigma}=-2\uppi^{2}\mu I_{0}\nu\dot\nu/(A+B)$ is the stress accumulation rate,
\begin{eqnarray}
\dot{\sigma}&\simeq&2.9\times 10^{18}M_{1.4}R_{10}^{2}A_{53}^{-1}\mu_{30}(\frac{\nu}{0.308~\rm{Hz}})
\nonumber\\
&&\times(\frac{\dot\nu}{-1.356\times 10^{-12}~\rm{Hz/s}})~\rm{erg~cm^{-3}~yr^{-1}},
\end{eqnarray}
$\mu_{30}=\mu/(10^{30}~\rm{erg~cm^{-3}})$.
In Eq.(\ref{eq15}), $|\Delta\sigma|$ is the stress relieved during glitches (larger than the stress accumulated between two successive glitches if the sequence of starquakes is history-dependent~\citep{2020MNRAS.491.1064G}).
If glitches are triggered by starquakes, $|\Delta\sigma|\ll \sigma_{\rm c}$, otherwise, glitches due to starquakes in NSs should be much less frequent than observed~\citep{2020MNRAS.491.1064G,2021arXiv210612604R}.
Using the shear modulus $\mu=10^{32}~\rm{erg/cm^{3}}$, the stress relieved during starquake is estimated to be $10^{18-24}~\rm{erg/cm^{3}}$~\citep{2004APh....22...73Z}, which means that $|\Delta\sigma|\simeq10^{16-22}~\rm{erg/cm^{3}}$ for a NS with $\mu\sim10^{30}~\rm{erg/cm^{3}}$ (because $\dot\sigma \propto \mu$).
The corresponding waiting time will be
\begin{equation}
\label{eq16}
\Delta t_{q}\simeq (0.0035-17500)~\rm{yr}
\end{equation}
assuming the NS has a mass $M=1.4~M_{\odot}$, radius $R=10~\rm{km}$ and $A_{53}=0.2-1$.
According to Eqs.(\ref{eq9}) and (\ref{eq12}), for the first case, glitch size of the starquake will be
\begin{equation}
\label{eq17}
\frac{\Delta\nu}{\nu}\simeq \frac{B}{A+B}|\dot\varepsilon|\Delta t_{q}\simeq 4.5\times 10^{-18}-1.1\times 10^{-16},
\end{equation}
while for the second one,
\begin{equation}
\label{eq18}
\frac{\Delta\nu}{\nu}\simeq 1.4\times 10^{-21}-7.1\times 10^{-15}.
\end{equation}
According to Eqs.(\ref{eq1}) and (\ref{eq9}), the elastic energy release will be
\begin{eqnarray}
\label{starquake energy}
\Delta E&\simeq& 2B\varepsilon_{0}(\Delta\varepsilon_{0}-\Delta\varepsilon)=2A\varepsilon_{0}\Delta\varepsilon=2A\varepsilon_{0}\Delta\nu/\nu
\nonumber\\
&\simeq& (3\times 10^{29}-1.6\times 10^{36})M_{1.4}R_{10}^{2}\nu_{100}^{2}~\rm{erg},
\end{eqnarray}
which is at least four orders of magnitude lower than $E_{\rm iso-X}$.

Apart from this kind of long-term spin down strained starquakes, pulsars with long spin periods could occasionally suffer mini contractions due to lose of centrifugal force~\citep{2020arXiv200708370D}, resulting in decrease in moment of inertia and release of huge amount of gravitational energy.
This process is absolutely the same with the so-called type II glitches or bulk-variable starquakes in strangeon stars~\citep{2014MNRAS.443.2705Z}.
According to \cite{2014MNRAS.443.2705Z}, the glitch size is expressed as
\begin{equation}
\frac{\Delta\nu}{\nu}=-\frac{\Delta I}{I}=-\frac{2\Delta R}{R},
\end{equation}
where $\Delta R (<0)$ is the radius decrease.
The energy release is
\begin{equation}
\Delta E=\frac{3GM^{2}}{10R}\frac{\Delta\nu}{\nu}=1.55\times 10^{53}M_{1.4}^{2}R_{10}^{-1}(\frac{\Delta\nu}{\nu})~\rm{erg}.
\end{equation}
If FRB 200428 is powered by this kind of mini contractions, assuming the majority of gravitational energy is emitted through X-ray emission, the corresponding glitch size will be
\begin{equation}
\label{eq19}
\frac{\Delta \nu}{\nu}\simeq \frac{E_{\rm iso-X}}{3GM^{2}/10R}=7.4\times 10^{-14}D_{10}^{2}M_{1.4}^{-2}R_{10}.
\end{equation}
The corresponding glitch is still much too tiny to be detectable.

Note that, starquake may trigger vortex unpinning which will enlarge the glitch size~\citep{2000ASSL..254...95E,2020MNRAS.499..455L}.
However, as the strain energy is much less than $E_{\rm iso-X}$, there is no need to discuss thermal effect of starquake in NSs.

We conclude that, just as expected, starquake of NSs will only result in pretty tiny glitches with sizes of $1.4\times10^{-21}-7.1\times10^{-15}$, which is far from detectable.
Besides, for the long-term stress-accumulation-type glitches, the elastic energy released during starquake is far from sufficient to explain the energy release during the X-ray burst if the release of magnetic energy in the magnetosphere or that in the crust is not taken into account.
If a glitch is found to be associated with FRB 200428, the spin down strained starquake scenario of NSs will be disfavored, magnetic energy should be included to explain the high energy counterpart under the framework of starquake of NSs.

\subsection{Spin down strained Type I and Type II glitches in strangeon stars (SSs)}
Strangeon star is a kind of solid quark star with no free magnetic energy in the magnetosphere or in the crust, therefore, starquakes in SSs are gravitationally and elastically powered~\citep{2003ApJ...596L..59X}.

We discuss type I glitches in SSs first.
Both type I glitches in SSs and the starquake process in NSs result from stress accumulation
during secular spin down phase, the differences lie in two aspects, the volume of the solid crustal component and the shear modulus.
Compared with NSs, SSs are totally solid, besides, a statistical research on glitch activity of SSs indicates $B\simeq A$, which means that the average shear modulus could be as high as $3\times10^{34}~\rm{erg/cm^{3}}$ ~\citep{2021MNRAS.500.5336W} for a SS with mass $M=1.4~M_{\odot}$ and radius $R=10~\rm{km}$.
These two aspects make it possible for SSs to produce large glitches.
Note that, the strangeon star is totally solid and assumed to be incompressible, just like that in~\cite{1971AnPhy..66..816B}.
Therefore, parameter $A$ in Eq.(2) will be used for strangeon stars, according to Eq.(\ref{eq3}), $A_{53}=0.621M_{1.4}^{2}R_{10}^{-1}$ in this case.

The size of type I glitch can be expressed as the following according to Eqs.(\ref{eq9}) and (\ref{eq11}),
\begin{equation}
\label{eq20}
\frac{\Delta\nu}{\nu}=\frac{(k-1)\Delta\varepsilon}{Q}=\frac{(k-1)B\Delta\varepsilon_{0}}{(A+B)Q}\simeq \frac{(k-1)|\dot\varepsilon|\Delta t_{q}}{Q},
\end{equation}
$k\sim 5-10$ is a constant presented above, $Q$ is the recovery coefficient, $Q\simeq 10^{-3}-1$ for the whole group of glitches.
However, $Q$ is generally close to $1$ for glitches in magnetars and high magnetic field pulsars, for example, $Q=1.0$ for the glitch on MJD 56756.0 in SGR J1822-1606, $Q=0.63(5)$ for the glitch on MJD 52464.01 in 1E 1841-045, $Q=0.185(10)$ for the glitch on MJD 52443.13(9) in 1E 2259+586, $Q=0.84(3)$ and $Q=0.81(4)$ for glitches on MJD 53290 and MJD 54240 in the high magnetic field pulsar J1119-6127~\footnote{https://www.atnf.csiro.au/people/pulsar/psrcat/glitchTbl.html}.

The waiting time $\Delta t_{q}$ can be estimated in similar ways with the starquake scenario in NSs.
From the mean glitching rate versus the spin down rate relation, $\Delta t_{q}\simeq 11~\rm{yr}$, which is the same with that in NSs.
From the view of stress accumulation, the relieved stress during glitches will be $\Delta\sigma\simeq3\times 10^{20-26}~\rm{erg/cm^{3}}$ if $\mu\simeq 3\times 10^{34}~\rm{erg/cm^{3}}$.
The increase of relieved stress during glitches has no impact on the estimated waiting time since $\dot\sigma \propto \mu$.
Therefore, $\Delta t_{q}\simeq (0.0035-17500)~\rm{yr}$ in this case, the same with that presented in Eq.(\ref{eq16}).

If $\Delta t_{q}\simeq 11~\rm{yr}$, according to Eqs.(\ref{eq12}) and (\ref{eq20}), the corresponding glitch size will be
\begin{eqnarray}
\label{eq21}
\frac{\Delta\nu}{\nu}&\simeq &(1-2.3)\times 10^{-9}M_{1.4}^{-1}R_{10}^{3}(\frac{0.1}{Q})(\frac{\nu}{0.308~\rm{Hz}})
\nonumber\\
&&\times (\frac{\dot\nu}{-1.356\times 10^{-12}~\rm{Hz/s}}),
\end{eqnarray}
note that $A_{53}=0.621M_{1.4}^{2}R_{10}^{-1}$ in the strangeon star case.
If we expect a glitch of size $\sim 10^{-6}$ in SGR 1935+2154, we should expect that this glitch has a recovery coefficient $Q\sim 10^{-4}$.
On the other hand, if the starquake accounts for the FRB, the total energy release during the glitch should exceed $E_{\rm iso-X}\simeq 1.148\times 10^{40}D_{10}^{2}~\rm erg$.
According to Eq.(54) in \cite{2021MNRAS.500.5336W}, the total energy release during type I glitches in SSs includes gravitational and elastic energy, if $A\simeq B$, the total energy release for SGR J1935+2154 will be
\begin{equation}
\label{eq22}
\Delta E\simeq2A\varepsilon_{0}\Delta\varepsilon= 2A\varepsilon_{0}|\dot\varepsilon|\Delta t_{q}\simeq5.6\times 10^{39}R_{10}^{5}\nu_{100}^{2}~\rm{erg},
\end{equation}
which is very close to $E_{\rm iso-X}$ assuming a distance of $10~\rm{kpc}$.
The corresponding burst rate is $1/\Delta t_{q}\sim 0.09~\rm{yr^{-1}}$, consistent with the burst rate of $(0.007-0.04)~\rm{yr^{-1}}$~\citep{2020Natur.587...54C}.

If $\Delta t_{q}\simeq (0.0035-17500)~\rm{yr}$, according to Eqs.(\ref{eq12}) and (\ref{eq20}), the corresponding glitch size will be
\begin{eqnarray}
\label{eq23}
\frac{\Delta\nu}{\nu}&\simeq &(3.2\times 10^{-14}-3.6\times 10^{-6})M_{1.4}^{-1}R_{10}^{3}(\frac{0.1}{Q})
\nonumber\\
&&\times (\frac{\nu}{0.308~\rm{Hz}})(\frac{\dot\nu}{-1.356\times 10^{-12}~\rm{Hz/s}}).
\end{eqnarray}
Upper limit of the glitch size could be as large as $3.6\times 10^{-6}$ even if $Q\sim 0.1-1$, however, the corresponding burst rate is $1/\Delta t_{q}\sim 5.7\times 10^{-5}~\rm{yr^{-1}}$, more than two orders of magnitude lower than the lower limit of estimated burst rate $0.007~\rm{yr^{-1}}$.
Similar with Eq.(\ref{eq22}), the total energy release will be
\begin{equation}
\label{eq24}
\Delta E \simeq 2A\varepsilon_{0}|\dot\varepsilon|\Delta t_{q}
\simeq(1.8\times10^{36}-9\times10^{42})R_{10}^{5}\nu_{100}^{2}~\rm{erg}.
\end{equation}
It seems that, the energy requirement could be fulfilled in this case.
However, this energy release is inconsistent with observations, as such a large energy release have never been observed to be accompanied by glitches in the Vela pulsar.
Therefore, the long waiting time case can be ruled out from the view of both burst rate and energy release.

Type II glitches in SSs arise from accretion or loss of centrifugal force.
Similar with the mini contraction picture in NSs, huge amount of gravitational energy will be released.
Therefore, the energy release during the X-ray burst gives constraint on the size of the type II glitch.
The glitch size has been presented in Eq.(\ref{eq19}).

To sum up, it is possible for SSs to generate detectable type I glitches of size $10^{-6}$ if the recovery coefficient is $Q\sim10^{-4}$.
A waiting time of about $11~\rm{yr}$ will also fulfill the burst rate estimate.
Note that, it is possible that local mini contractions are accompanied by type I glitches.
If this happens, the tension between energy requirement and supply could be eased for the case of $\Delta t_{q}\simeq 11~\rm{yr}$.
On the other hand, if $Q$ is similar with those observed in other magnetars and high magnetic field pulsars, i.e., $Q\sim 0.1-1$, a waiting time of more than $2000~\rm{yr}$ will be needed to produce glitches of size $10^{-6}$, but such a long waiting time is inconsistent with the burst rate.
We stress that, the recovery coefficient is crucial in distinguishing glitch mechanisms of NSs and SSs if glitches are detected to be accompanied by FRBs.

\subsection{Magnetically strained local starquakes in magnetars}

In the standard magnetar model, magnetars are young, isolated neutron stars powered by magnetic energy, with crustal magnetic fields which are both strong ($B>10^{14}-10^{15}~\rm{G}$) and multipolar~\citep{1995MNRAS.275..255T,1996ApJ...473..322T,2002ApJ...574..332T},
the ellipticity could be as high as $10^{-4}$, depending on the equation of state and magnetic field configurations~\citep{2008MNRAS.385..531H}.
Within the magnetic decay timescales of $\sim 10^{5}-10^{7}~{\rm{yr}}$~\citep{2009A&A...496..207P,2013MNRAS.434..123V,2016MNRAS.456...55G,2019AN....340.1030G} through mechanisms such as Hall drift and Ohmic dissipation, magnetic stress develops in the crust due to internal magnetic field strength evolution and magnetic field reconfiguration~\citep{1998ApJ...492..267R}.
Once the magnetic stresses exceeds the critical threshold that the crust can sustain (the strain reaches the critical strain angle, $\theta_{{\rm{cr}}}$), the crust cracks locally (named crust fractures)~\citep{2010MNRAS.407L..54C,2018MNRAS.480.5511B}.
This process has been supposed to account for burst properties of FRBs~\citep{2019MNRAS.488.5887S} and will also result in glitches~\citep{1991ApJ...382..587R,1998ApJ...492..267R,2015MNRAS.449.2047L}.

If FRBs originate from crust fracture processes in young magneatrs, glitches may be accompanied by the crust fractures.
We present order of magnitude estimation of size of the glitch that may be associated with FRB 200428 in this subsection.
According to~\cite{2015MNRAS.449.2047L}, crust fractures tend to occur in the outer equatorial region first.
Size of the glitch induced by crust fracture contains two parts, the pure starquake resulting from moment of inertia decrease, and glitch size amplified by sudden unpinning of vortex lines accompanied by crustal movement, among which the latter dominates~\citep{2018MNRAS.473..621A}.
We estimate glitch size contributed by moment of inertia decrease here and left the corresponding vortex unpinning to be discussed in subsection 4.2.

To estimate the moment of inertia decrease, volume and surface area of the broken region should be known first.
Lander et al. found that, the magnetic energy release is independent of the field strength~\citep{2015MNRAS.449.2047L}, the energy release is related to the broken region through
\begin{equation}
\label{burst magnetic energy}
E_{\rm{iso-X}}\simeq 2.3\times 10^{28}(\frac{\theta_{{\rm{cr}}}}{0.01})d^{2}l~\rm{erg},
\end{equation}
where $0.01\lesssim \theta_{{\rm{cr}}}\lesssim 0.1$~\citep{2009PhRvL.102s1102H,2010MNRAS.407L..54C,2012MNRAS.426.2404H,2018MNRAS.480.5511B}, $d$ is the fracture depth and $l$ is the fracture length,  $d^{2}l$ represents the effective volume of the broken region.
The fracture geometry is complicated, depending on the topological structure of the magnetic field~\citep{2000ApJ...543..987F,2015MNRAS.449.2047L}.
From Eq.(\ref{burst magnetic energy}), it is clear that, for a certain amount of energy release, $d^{2}l$ is a function of $\theta_{\rm cr}$.
The maximum glitch size occurs when the broken region moves from the equatorial region to the polar region, therefore,
\begin{eqnarray}
\frac{\Delta \nu}{\nu}&=&-\frac{\Delta I}{I}\leq\frac{\rho_{\rm c}(d^{2}l)R^{2}}{2MR^{2}/5}\sim 4\times 10^{-9}(\frac{E_{\rm{iso-X}}}{10^{40}~\rm{erg}})
\nonumber\\
&&\times(\frac{\theta_{\rm cr}}{0.01})^{-1}M_{1.4}^{-1}\rho_{c-13},
\end{eqnarray}
where $\rho_{\rm c}$ is the crustal density, $\rho_{{\rm{c}}-13}=\rho_{\rm c}/(10^{13}~\rm{g})$.
However, the fracture displacement is much small than the radius, by a factor of $\sim 10^{-4}$~\citep{2018MNRAS.473..621A}, the glitch size should be lowered by a factor of about $\sim 10^{-4}$ simultaneously.
Therefore, considering the uncertainty of $\theta_{\rm cr}$, the glitch size will be
\begin{equation}
4.5\times 10^{-14}<\frac{\Delta\nu}{\nu}<4.5\times 10^{-13},
\end{equation}
which is too small to be detected.
Contribution of vortex unpinning induced by crustal movement is further discussed in subsection 4.2.

\section{Orbital angular momentum transfer scenario}

Collisions between NSs and small celestial bodies such as asteroids or comets are also promising to produce FRBs and induce glitches simultaneously.
Asteroids/comets are small celestial bodies with typical mass of $10^{18}~\rm{g}$~\citep{2008ApJ...682.1152C}.
Previous literature has suggested the possibility of collision between NSs and asteroids/comets~\citep{1981ApJ...248..771C,1990Natur.344..313M,1994ApJ...437..727K}.
Huang et al. argued that, collisions between NSs and asteroids/comets could reasonably explain the main FRBs features such as their durations and radio luminosities~\citep{2015ApJ...809...24G}.
Besides, asteroids/comets will bring orbital angular momentum onto NSs, resulting in sudden glitches or anti-glitches~\citep{2014ApJ...782L..20H}.

The glitch size depends on mass of the asteroid/comet, which could be constrained by the energy release of the X-ray burst associated with FRB 200428.
To estimate the mass and the maximum glitch size reachable, two extreme cases are considered here.
Firstly, the asteroid/comet triggers magnetic energy release through some unknown ways during the impact process, in this case, the X-ray burst may have nothing to do with mass of the asteroid/comet.
Therefore, we use the typical mass of $10^{18}~\rm{g}$ for calculations.
This mass represents the lower mass limit of the asteroid/comet.
Secondly, the energy release of the X-ray burst comes from the gravitational energy of the asteroid/comet, this mass sets the upper mass limit of the asteroid/comet, it can be estimated through
\begin{equation}
E_{\rm iso-X}=\eta\frac{GMm}{R},
\end{equation}
where $m$ is the mass of the asteroid/comet, $\eta$ is the coefficient converting gravitational energy into X-ray emission, typically, $\eta\simeq (0.01-1)$.
In this case, the maximum mass allowed will be
\begin{equation}
m=5.35\times 10^{21}(\frac{E_{\rm{iso-X}}}{10^{40}~\rm{erg}})(\frac{\eta}{0.01})^{-1}M_{1.4}^{-1}R_{10}~\rm{g}.
\end{equation}
Therefore, approximate mass of the asteroid/comet lies in the range $10^{18}~\rm{g}$ to $6.2\times 10^{21}~\rm{g}$ assuming the central pulsar has a mass $M=1.4~M_{\odot}$ and radius $R=10~\rm{km}$.
Note that, the maximum mass is proportional to $E_{\rm iso-X}$, it could be overestimated as the X-ray emission could be anisotropic.

The maximum glitch size is estimated in the following.
According to Eq.(1) in \cite{2014ApJ...782L..20H}, the asteroid/comet contributes the most angular momentum if the periastron of its orbit falls exactly on the surface of the NS, in this case, the orbital angular momentum is $m(2GM/R)^{1/2}R$.
The maximum glitch size can be estimated through conservation of angular momentum,
\begin{equation}
I\cdot 2\uppi\nu\pm m(2GM/R)^{1/2}R=I'\cdot 2\uppi \nu',
\end{equation}
\begin{equation}
I'=\frac{2}{5}(M+m)R^{2}\simeq \frac{2}{5}MR^{2}=I,
\end{equation}
$\nu'$ is the spin frequency after the collision, therefore,
\begin{eqnarray}
\frac{\Delta\nu}{\nu}&=&\frac{\nu'-\nu}{\nu}\leq \pm \frac{mR(2GM/R)^{1/2}}{2\uppi \nu I}\sim \pm9\times 10^{-12}
\nonumber\\
&&\times(\frac{m}{10^{18}~{\rm{g}}})M_{1.4}^{-1/2}R_{10}^{-3/2}.
\end{eqnarray}

For this section, we conclude that, in the most optimistic case, size of the sudden glitch/anti-glitch induced by orbital angular momentum transfer shall be in the range $9\times 10^{-12}-5.6\times 10^{-8}$.
If mass of the asteroid is $\sim 10^{20}~\rm{g}$ as proposed by Dai~\citep{2020ApJ...897L..40D},
the maximum energy deposition will be \begin{equation}
\label{e-dep}
\Delta E_{\rm dep} \sim mc^{2}\sim 9\times 10^{40}~\rm{erg},
\end{equation}
the corresponding glitch/anti-glitch size will be $\sim 0.9\times 10^{-9}$.
If the pulsar is a NS rather than a SS, collisions may also trigger vortex unpinning, this possibility will be discussed in subsections 4.1 and 4.3.

\section{Vortex unpinning scenario}

Vortex unpinning is another efficient way to transfer angular momentum and result in glitches~\citep{1969Natur.224..673B,1975Natur.256...25A}.
In this section, we discuss glitch size contributed by vortex unpinning induced by starquakes, collisions between NSs and asteroids/comets, and crust fractures.

Since we are talking about vortex unpinning, one critical point should be discussed first, are there superfluid neutrons in the crust of young magnetars?
Nobody knows the exact answer, but it can be responded from three aspects.
Firstly, a recent statistical work found a constant ratio $\dot\nu_{\rm g}/|\dot\nu|=0.010\pm 0.001$ for all rotation powered pulsars, high magnetic field pulsars and magnetars, reflecting their similar fractional moment of inertia (FMoI) of crustal superfluid neutrons~\citep{1999PhRvL..83.3362L,2017A&A...608A.131F}.
$\dot\nu_{\rm g}$ is the average glitch activity, defined in Equation (1) in ~\cite{2017A&A...608A.131F}.
Secondly, magneto-thermal evolution simulations suggest that, although magnetars exhibit high surface temperature and thermal luminosity, there exists a sharp decrease in temperature in the density range $10^{11}-10^{13}~\rm{g/cm^{3}}$~\citep{2009MNRAS.395.2257K}, making it possible for at least part of the crust to be superfluid.
Thirdly, the anomalous X-ray pulsar (AXP), 1E 1841-045, a glitcher with a surface temperature of about $T_{\rm{s}}\sim 3\times 10^{6}~\rm{K}$~\citep{2013MNRAS.434..123V}, shares many similarities with SGR J1935+2154 from the aspects of high surface temperature, characteristic age and spin-inferred dipolar magnetic field (see Table \ref{AXP 1E 1841-045 tabel} in this paper).
Inferred FMoI for 1E 1841-045 is $1.2\pm 0.2$ percentage, which is a little bit smaller than other mature pulsars with larger characteristic ages and lower surface temperature (see Table 1 in~\cite{2015SciA....1E0578H} for comparison).
From all these considerations, it is reasonable to speculate that superfluid neutrons do exist in the crust of SGR J1935+2154, just like that in 1E 1841-045.

\begin{table}
\begin{center}

    \caption{Parameters of AXP 1E 1841-045}
    \begin{tabular}{ | p{1.7cm} | p{3.0cm} |p{2.0cm}|}
    \hline
    Parameter          & Value                                & Refs.$^b$\\ \hline
    $\nu$              & $0.085~\rm{Hz}$                      & (1)\\
    $\dot\nu$          & $-2.95\times 10^{-13}~\rm{Hz/s}$     & (1)\\
    $B_{d}$            & $7\times 10^{14}~\rm{G}$             & (1)\\
    $L_{\rm{sd}}$      & $\sim 10^{33}~\rm{erg/s}$            & (2)\\
    $\tau_{\rm c}$     & $4.7~\rm{kyr}$                       & (2)\\
    $\tau_{\rm k}$     & $(0.75-2.1)~\rm{kyr}$                  & (2)\\
    $T_{\rm s}$        & $\sim 3\times 10^{6}~\rm{K}$         & (3)\\
    \hline
    \end{tabular}
    \label{AXP 1E 1841-045 tabel}
\end{center}
$^b$ \scriptsize{References: (1)~\cite{2002ASPC..271..309G};(2)~\cite{2014ApJ...781...41K};(3)~\cite{2013MNRAS.434..123V}.} \\
\end{table}

In the following calculations, we assume SGR J1935+2154 has a FMoI of crustal superfluid of $\sim 1.2$ percentage, comparable with that in 1E 1841-045.
Surface temperature of SGR J1935+2154 is unknown, but it is very possible that SGR J1935+2154 is a Crab-like pulsar with high surface temperature given the magnetic energy release in the crust.
By Crab-like (Vela-like), we mean pulsars with surface temperature and characteristic ages similar to those of the Crab pulsar (Vela pulsar).

According to~\cite{1984ApJ...276..325A}, pinning arises due to the interaction between vortex and nucleus, it is energetically favorable if the cost per particle to form its core is reduced, in other words, the total energy of the vortex-nuclei system is lowered.
The energy cost to pin for a vortex line is $\Delta^{2}/E_{\rm f}$~\citep{1984ApJ...276..325A}, where $\Delta$ is the energy gap of superfluid neutrons, $E_{\rm f}$ is the Fermi energy of neutrons.
The pinning energy is $E_{\rm p}=\frac{3}{8}\gamma (\frac{\Delta^{2}}{E_{\rm f}}n_{\rm f})V_{\rm vn}$, where $\gamma$ is close to unity, $n_{\rm f}=\frac{k_{\rm F}^{3}}{3\uppi^{2}}$ is number density of bulk neutrons, $k_{\rm F}$ is the Fermi momentum of superfluid neutrons, $V_{\rm vn}=\frac{4}{3}\uppi \xi^{3}$ is the overlap volume between vortex and the pinned nucleus,
$\xi\simeq 10~\rm{fm}$ is the radius of coherence length of bulk neutrons.
Essentially, vortex unpins when the angular velocity difference between the crust (and that coupled to it) and the superfluid component exceeds the critical lag~\citep{1984ApJ...276..325A}.

The glitch size may be amplified by vortex unpinning.
As stated above, the starquake process is accompanied by strain energy release, the collision process will deposit gravitational energy as thermal and kinetic energies, both of which may induce vortex unpinning.
Besides, crust fracture itself may act as a glitch triggering mechanism~\citep{1991ApJ...382..587R,1998ApJ...492..267R}.
We discuss the thermal effect of starquakes and collisions between NSs and asteroids/comets in subsection 4.1, vortex unpinning induced by crust fracture is discussed in subsection 4.2, neutron scattering induced by collisions is discussed in subsection 4.3.
The corresponding glitch sizes are estimated respectively.

\subsection{Thermally driven glitches}

Link et al. studied the thermal and dynamical responses of a NS to a sudden internal heating in the inner crust for Crab-like, Vela-like and older pulsars~\citep{1996ApJ...457..844L}.
The local temperature increase affects the coupling between superfluid neutrons and the outer crust, resulting in the spin-up phenomenon.
The relation between energy deposition and glitch size is presented in Fig. 15 in their work.
They found that energy deposition of $\sim10^{40}~\rm{erg}$ produces delayed spin-ups (also named slow rise glitches) with size just above timing noise level for three kinds of pulsars, while energy deposition of $\sim 2.1\times 10^{42}~\rm{erg}$ in the crust of hot Crab-like pulsars (surface temperature of the Crab pulsar is about $1.6\times 10^{6}~\rm{K}$~\citep{1985ApJ...288..191A}) produces glitches of size $\Delta\nu/\nu~\sim 7\times 10^{-8}$ in a timescale of one day~\citep{1996ApJ...457..844L}.
Energy deposition of $\sim 10^{42}~\rm{erg}$ in the crust of Vela-like pulsars produces glitches of size $\Delta\nu/\nu~\sim 10^{-6}$ within a timescale of minutes.
Comparisons between the energy deposited by NS starquakes, collisions between NSs and asteroids/comets and the energy used by Link et al. may shed light upon the corresponding glitch sizes in our cases.

For the NS starquake case, according to Eq.(\ref{starquake energy}),
the elastic energy release is $\Delta E\simeq (6.6\times 10^{26}-7.2\times 10^{32})~\rm{erg}\ll 10^{40}~\rm{erg}$.
This amount of energy generates glitches well below timing noise level no matter SGR J1935+2154 is Crab-like, Vela-like or even older.
For the collision case, mass of the asteroid/comet is constrained to be $(10^{18}-6.2\times 10^{21})~\rm{g}$ in section 3, which corresponds to the maximum energy deposition of $\Delta E_{\rm grav}\simeq (9\times 10^{38}-5.6\times 10^{42})~\rm{erg}$ on the surface of the NS.
This energy comes from the gravitational energy of the asteroid/comet, and was converted into thermal and kinetic energy.
We discuss effect of the thermal energy in this subsection and left that of kinetic energy to be discussed in subsection 4.3.

The thermal energy of collisions is unlikely to be important in the case of SGR J1935+2154.
This is because, crustal temperature is higher than the surface temperature, most of the thermal energy deposited by collisions tends to be radiated through thermal emission rather than be transferred inward into the crusts.
Besides, even if part of thermal energy is transferred into the crust, timescale of this process is comparable to thermal relaxation timescale of the crust, which is at least tens of days~\citep{2015ApJ...809L..31D}.
Therefore, this process will induce slow rise glitch rather than sudden glitch, which is hard to be detected for magnatars (outbursts appear frequently).
Moreover, thermal energy deposited will be no larger than the gravitational energy, which is about $(9\times 10^{38}-5.6\times 10^{42})~\rm{erg}$ as stated above.
Given the heat conduction timescale and efficiency, size of this slow rise glitch may not be larger than $7\times 10^{-8}$ if SGR J1935+2154 is Crab-like or $< 10^{-6}$ if it is Vela-like, as presented by \cite{1996ApJ...457..844L}.
As spin frequency of SGR J1935+2154 is low compared with the Crab and Vela pulsars (the Crab and Vela pulsars have spin frequency $\nu=29.612~\rm{Hz}$ and $\nu=11.198~\rm{Hz}$ separately), the glitch sizes may be amplified at most by the frequency ratios, which are $\sim 96$ and $\sim 36$ respectively according to Eq.(21) in~\cite{1996ApJ...457..844L}.
The corresponding glitch size is $< 6.7\times 10^{-6}$ for a Crab-like pulsar or $<3.6\times 10^{-5}$ for a Vela-like pulsar.
If mass of the asteroid/comet is reduced to $\sim 10^{20}~\rm{g}$~~\citep{2020ApJ...897L..40D}, the maximum energy deposition reduces to $\Delta E_{\rm dep}\sim 9\times 10^{40}~\rm{erg}$, therefore, the corresponding maximum glitch size for SGR J1935+2154 reduces to $<10^{-7}$ (Crab-like) or $< 10^{-6}$ (Vela-like) according to Fig.15 in~\cite{1996ApJ...457..844L}.

Note that, the above results are only order of magnitude estimations, simulations are needed to comprehensively estimate size of the thermally driven glitch, parameters such as surface and crustal temperature of SGR J1935+2154 are therefore urgently needed.

\subsection{Vortex unpinning induced by crust fractures}
As stated in subsection 2.3, crust fracture may amplify glitch size through vortex unpinning.
The glitch size depends on the number of vortices attached to the broken crustal plate, vortices unpinned through crustal movement can induce vortex avalanche by perturbing nearby vortices which are close to unpin if the pulsar is Vela-like rather than Crab-like~\citep{1993ApJ...409..345A}.
However, the surface area and the fracture geometry of the broken region depend on the topological structure of the magnetic field~\citep{2000ApJ...543..987F,2015MNRAS.449.2047L}, which is hard to be known except simulation.
For order of magnitude estimation, we assume the broken plate is cubic and calculate the number of vortices that can be unpinned.
Note that, this method has been used in~\cite{2018MNRAS.473..621A}, they have shown that the number of vortices only differs by a factor of several compared with a prism broken plate.
An illustration of the crust fracture is shown in Fig.1.

According to Eq.(\ref{burst magnetic energy}), the effective volume of the broken region is
\begin{equation}
V_{\rm eff}=d^{2}l\simeq 4.3\times 10^{11}(\frac{E_{\rm{iso-X}}}{10^{40}~\rm{erg}})(\frac{\theta_{\rm cr}}{0.01})^{-1}~{\rm{cm^{3}}}.
\end{equation}
Taking length of the cubic plate to be $\Delta l$, $E_{\rm{iso-X}}\sim 10^{40}~\rm{erg}$, therefore,
\begin{equation}
\Delta l\simeq(35-75)~\rm{m},
\end{equation}
the upper limit corresponds to $\theta_{\rm cr}\sim 0.01$.
Previously, post-glitch timing fits by vortex creep model to the largest Crab glitch in 2017 and to the peculiar glitch of PSR J1119-6127 yielded estimates of sizes of the broken crustal plates to be $\Delta l\simeq (6-18)~\rm{m}$~\citep{2019MNRAS.488.2275G} and $\Delta l\simeq 6~\rm{m}$~\citep{Akbal2015}.
Similarly, timing fits to the smallest glitch in the Crab pulsar yielded $\Delta l\sim 160~\rm{m}$ or $\Delta l\sim 75~\rm{m}$, depending on geometrical assumptions of the broken plates~\citep{2018MNRAS.473..621A}.
To sum up, our estimates are in good agreement with previous results.

The number of unpinned vortices, $N$, triggered by crustal movement, is
\begin{equation}
\label{vortex number1}
N\simeq (\Delta l)^{2}n_{\rm v}\simeq 2.4\times10^{10}-1.1\times 10^{11},
\end{equation}
where $n_{\rm{v}}=2\Omega/\kappa=6.35\times 10^{3}(\nu/\rm Hz)~\rm{cm^{-2}}$ is the surface density of vortices, $\Omega=2\uppi\nu$, $\kappa=h/(2m_{n})$ is the vorticity quantum, $m_{n}$ is the mass of neutrons, $h$ is the Planck constant.
The change in superfluid rotation rate, $\delta\Omega$, is related to the number of unpinned vortices through
\begin{equation}
\label{spin lag}
N=\frac{2\uppi R^{2}\delta\Omega}{\kappa}.
\end{equation}
Finally, based on Eqs.(\ref{vortex number1}), (\ref{spin lag}) and angular momentum conservation~\citep{1993ApJ...409..345A}, the glitch size will be
\begin{equation}
\frac{\Delta\nu}{\nu}\sim \frac{I_{\rm s}}{I_{\rm c}}\frac{\delta \Omega}{\Omega}\sim 4\times 10^{-8}-1.8\times 10^{-7},
\end{equation}
where $I_{\rm s}$ is the moment of inertia of crustal superfluid, $I_{\rm c}$ is the total moment of inertia that coupled to the crust, $I_{\rm c}\simeq I$.
For order of magnitude estimation, $I_{\rm s}/I_{\rm c}\sim 10^{-2}$ is reasonable regardless of the crustal structural difference between Crab-like young pulsars and Vela-like mature pulsars~\citep{1993ApJ...409..345A}.

The glitch size can be further amplified considering the avalanche.
The crustal structural difference between Crab-like pulsars and Vela-like pulsars lies in that, the network of vortex unpinning region is sufficiently connected in Vela-like mature pulsars while not in Crab-like young pulsars.
Therefore, if the kinetic age of SGR J1935+2154 is reliable, SGR J1935+2154 tends to be Vela-like, larger glitches ($\Delta\nu/\nu>0.9\times 10^{-6}$) can be produced as avalanches.

\begin{figure}
\centering
\includegraphics[width=0.53\textwidth]{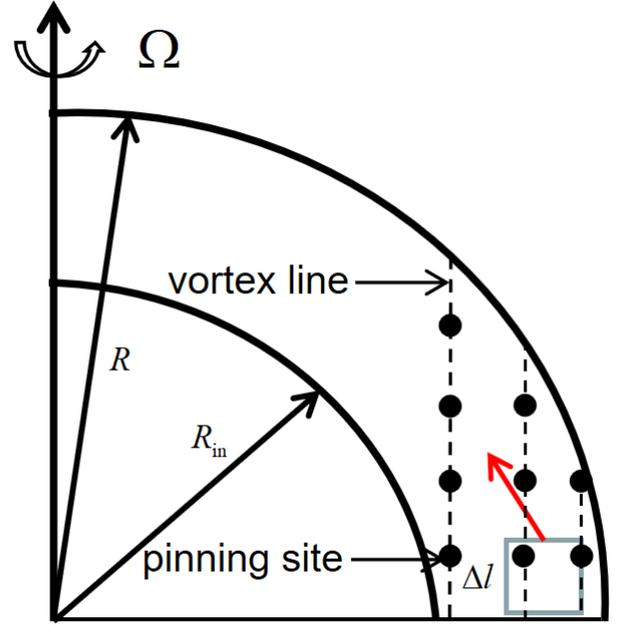}
\caption{An illustration of the crust fracture.
We take the geometry of the broken region to be a cubic plate with length $\Delta l$.
$R_{\rm in}$ is the radius of the core of the star, thickness of the crust is $R-R_{\rm in}$.
The red arrow represents motion direction of the cubic plate.
}
\label{figure 1}
\end{figure}

\subsection{Glitches induced by neutron-vortex scattering}

Different from the influence of thermal energy (by affecting the coupling between superfluid neutrons and the outer crust),
Layek \& Yadav proposed that, the strain energy released during starquake may excite unbound superfluid neutrons, inelastic scattering of these excited free neutrons with the vortex core neutrons nearby may overcome the pinning energy barrier and result in glitches of size $10^{-8}-10^{-7}$ for Crab-like pulsars and $10^{-9}-10^{-8}$ for Vela-like pulsars within a timescale of $(0.2-2)~\rm{s}$, the glitch size depends on the strain energy released~\citep{2020MNRAS.499..455L}.

The detailed picture proposed by~\cite{2020MNRAS.499..455L} is that, the NS breaks in the inner crust in the equatorial plane.
The strain energy ($\Delta E_{\rm strain}\sim 10^{40}~\rm{erg}$, or equally, $\Delta E_{\rm strain}\sim 6.24\times 10^{45}~\rm{MeV}$) is assumed to be isotropically distributed from the quake site, this energy makes free neutrons excited in a cubical region (similar with the cubic plate as shown in Fig. 1).
Scattering of these excited neutrons with the surrounding pinned vortices may further unpin a large amount of vortices and induce glitches.
Actually, neutrons could be thermally excited, the fractional excited neutrons in the superfluid state is $e^{-\frac{\Delta}{k_{B}T}}$, where $k_{B}$ is the Boltzmann constant, $T$ is the temperature.
Fraction of excited neutrons is generally small.
In ~\cite{2020MNRAS.499..455L}, neutrons are excited by absorption of elastic energy in the inner crust rather than thermally excited, ratio between excited neutrons to superfluid neutrons is set to be the constant $\Delta/E_{\rm f}\simeq 0.07$ for both Crab-like and Vela-like pulsars.

Actually, this kind of neutron excitation exists in thermal non-equilibrium state.
If neutrons can be excited by the sudden release of strain energy, they may also be excited by other sudden energy release processes, such as the kinetic energy release during collisions.
Both the strain and kinetic energy release represent the thermal non-equilibrium state.
Besides, the kinetic energy is released around the collision site, similar with the strain energy released around the crustquake site.
Therefore, it is reasonable to assume that, the kinetic energy released during collisions may also excite unbound superfluid neutrons and induce vortex unpinning.
In this case, calculations in~\cite{2020MNRAS.499..455L} applies here.
The upper limit of the kinetic energy is $\Delta E_{\rm k}\simeq (9\times 10^{38}-5.6\times 10^{42})~\rm{erg}$ for the mass range of $(10^{18}-6.2\times 10^{21})~\rm{g}$.
For order of magnitude analysis, we use the same parameters with ~\cite{2020MNRAS.499..455L} as it is very possible that SGR J1935+2154 is a Crab-like high temperature pulsar.
Therefore, $\Delta=0.06~\rm{MeV}$, $E_{\rm f}=0.83~\rm{MeV}$, Fermi momentum of superfluid neutron in the outer part of the inner crust is $k_{\rm F}\simeq 0.20~\rm{fm^{-1}}$~\citep{2008LRR....11...10C,2011PhRvC..84f5807P,2015PhRvC..91c5805S}.
Uncertainties of parameters will be briefly discussed at the end of this subsection.

Following the procedure in~\cite{2020MNRAS.499..455L}, we estimate the volume of the region ($V$) within which free neutrons could become excited first.
According to Eqs.(6) and (7) in their work, the volume $V$ can be expressed as
\begin{equation}
V=\frac{\Delta E_{\rm k}E_{\rm f}}{n_{\rm f}\Delta^{2}},
\end{equation}
where $E_{\rm f}>E_{\rm p}$.
$V$ is proportional to $\Delta E_{\rm k}$, therefore, $V$ will be a factor of $\Delta E_{\rm k}/\Delta E_{\rm strain}$ times the result presented by \cite{2020MNRAS.499..455L}, which is $5.1\times 10^{6}~\rm{m^{3}}$,
\begin{equation}
V=\frac{\Delta E_{\rm k}}{\Delta E_{\rm strain}}\times 5.1\times 10^{6}~\rm{m^{3}}\sim (4.6\times 10^{5}-2.9\times 10^{9})~\rm{m^{3}},
\end{equation}
set the length of each side of the cube to be $\Delta L$,
\begin{equation}
\Delta L=V^{1/3}\sim (77-1426)~\rm{m}.
\end{equation}
Number of excited neutrons is
\begin{equation}
N_{\rm ex}=\Delta E_{\rm k}/\Delta=9.4\times 10^{45}-5.8\times 10^{49}.
\end{equation}
Note that, the upper limit of $\Delta L$ is larger than the general thickness of the crust, which is $\Delta R\sim 1~\rm {km}$.
In this paper, we are confined to discuss within $\Delta L'\sim (77-1000)~\rm {m}$,
which corresponds to $V'\sim (4.6\times 10^{5}-10^{9})~\rm{m^{3}}$, $\Delta E_{\rm k}'\sim (9\times 10^{38}-1.9\times 10^{42})~\rm{erg}$, and $N_{\rm ex}'\sim 9.4\times 10^{45}-2\times 10^{49}$.
Note that, if collisions happen in the equatorial plane, the fractional decrease in pinning force per unit length is estimated similarly with Eq.(16) in~\cite{2020MNRAS.499..455L}, that is
\begin{equation}
\frac{\Delta f}{f}=\frac{\Delta L'}{L'}\simeq 0.06-0.22,
\end{equation}
where $f$ is the pinning force per unit length of vortex lines, $\Delta f$ is the decrease of pinning force per unit length of vortex lines due to vortex unpinning in volume $V'$, $L'\simeq (2R\Delta L')^{1/2}$ is the length of a vortex line.
This decrease of pinning force ensures that vortex lines will move under magnus force.

We calculate the number of vortices that can be unpinned.
Relation between the number of vortices which could be unpinned by excited neutrons $N_{\rm v}$ and the volume $V'$ is
\begin{equation}
V'=\frac{N_{\rm v} \Delta L'}{n_{\rm v}}.
\end{equation}
For SGR J1935+2154, $N_{\rm v}$ equals
\begin{eqnarray}
\label{vortex number}
N_{\rm v}&=&\frac{V'n_{\rm v}}{\Delta L'}=n_{\rm v}(V')^{2/3}=n_{\rm v}(\Delta E_{\rm k}')^{2/3}(\frac{E_{\rm f}}{n_{\rm f}\Delta_{\rm f}^{2}})^{2/3}
\nonumber\\
&\sim & (1.2\times 10^{11}-1.85\times 10^{13})(\frac{\nu}{0.308~\rm {Hz}}).
\end{eqnarray}
The total number of vortices in the crust is
$N_{\rm tot}=(2\uppi R\Delta R)n_{\rm v}\sim 1.2\times 10^{15}(\nu/0.308~{\rm Hz})(\Delta R/1~{\rm{km}})(R/10~\rm{km})$.
Finally, according to Eq.(4) in ~\cite{2020MNRAS.499..455L}, the glitch size will be
\begin{equation}
\frac{\Delta\nu}{\nu}=(\frac{I_{\rm s}}{I})(\frac{N_{\rm v}}{N_{\rm tot}})(\frac{2\Delta t_{q}}{\tau_{\rm c}})\sim (0.008-1.2)\times 10^{-6}(\frac{\Delta t_{q}}{11~\rm{yr}}),
\end{equation}
where $I_{\rm s}/I$ is the FMoI of crustal superfluid.
Note that, this wide range reflects its dependence on the kinetic energy, or the mass of the asteroid/comet essentially.
The upper limit could be larger, as the actual waiting time could be larger than the average value of $\Delta t_{q}\simeq 11~\rm{yr}$.
Vortices outside the cubical region could also unpin if SGR J1935+2154 is a Vela-like mature pulsar.

Note that, according to Eq.(14) in~\cite{2020MNRAS.499..455L}, the total pinning energy within the volume $V=5.1\times 10^{6}~\rm{m^{3}}$ is
\begin{equation}
E_{\rm tp}=3.1\times 10^{24}~{\rm{MeV}}\ll \Delta E_{\rm strain}\sim 6.24\times 10^{45}~\rm{MeV},
\end{equation}
obviously, the efficiency of transfer energy from excited neutrons to vortices is extremely low.
The main reason is that, the excited neutrons are confined within the volume just around the quake site due to their low mobility, and a large amount of excited neutrons have no vortices to attack.
For our case, $V'\sim (4.6\times 10^{5}-10^{9})~\rm{m^{3}}$, $\Delta E_{\rm k}'\sim (9\times 10^{38}-1.9\times 10^{42})~\rm{erg}$, the total pinning energy within $V'$ is
\begin{equation}
\label{Etp}
E_{\rm tp}'\sim (2.7\times 10^{23}-5.8\times 10^{26})~{\rm{MeV}}\ll \Delta E_{\rm k}',
\end{equation}
only an extremely small part of energy has been used to unpin vortices.

\begin{table*}
\begin{center}

    \caption{Summary of the glitch sizes for various physical processes and mechanisms.}
    \begin{tabular}{ | p{4.0cm} | p{4.0cm} |p{4.5cm}|}
    \hline\hline
    Physical process                     &mechanism          &$\Delta\nu/\nu$ $^c$\\
    \hline
    Global starquake in NSs              &(1)spin down strained starquake      &$\sim1.4\times 10^{-21}-7.1\times 10^{-15}$\\
                                         &(2)mini contraction                  &$\sim7.4\times 10^{-14}$\\
                                         &(3)vortex unpinning                  &neglectable\\
    \hline
    Global starquake in SSs              &(1)type I glitch                     &$\sim (1-2.3)\times 10^{-9}(0.1/Q)$\\
                                         &(2)type II glitch                    &$\sim 7.4\times 10^{-14}$\\
    \hline
    Crust fracture in magnetars          &(1)pure local starquake              &$\sim 4.5\times 10^{-14}-4.5\times 10^{-13}$\\
                                         &(2)vortex unpinning induced by crustal movement$^d$   &$\sim 4\times 10^{-8}-1.8\times 10^{-7}$\\
    \hline
    Collisions between NSs and asteroids/comets ($m\sim 10^{18}-6.2\times 10^{21}~\rm{g}$)  &(1)angular momentum transfer &$<5.6\times 10^{-8}$\\
                                         &(2)thermal driven$^e$                   &$<6.7\times 10^{-6}-3.6\times 10^{-5}$\\
                                         &(3)neutron-vortex scattering         &$\sim (0.008-1.2)\times 10^{-6}(\Delta t_{q}/11~\rm{yr})$\\
                                         \hline
    Collisions between NSs and asteroids/comets ($m\sim 10^{20}~\rm{g}$) &(1)angular momentum transfer &$\sim0.9\times 10^{-9}$\\
                                         &(2)thermal driven                   &$<10^{-7}-10^{-6}$\\
                                         &(3)neutron-vortex scattering         &$\sim 5.6\times 10^{-8}(\Delta t_{q}/11~\rm{yr})$\\
    \hline\hline
    \end{tabular}
    \label{glitch size summary}
\end{center}
$^c$ \scriptsize{Notes: These values are calculated assuming the pulsar has a mass $M=1.4~M_{\odot}$ and radius $R=10~\rm{km}$.} \\
$^d$ \scriptsize{Notes: The upper and lower limits correspond to the critical strain angle of $\theta_{\rm cr}\simeq 0.01$ and $\theta_{\rm cr}\simeq 0.1$ respectively.}\\
$^e$ \scriptsize{Notes: The upper and lower limits correspond to assumptions that SGR J1935+2154 is a Vela-like and a Crab-like pulsar respectively.}\\
\end{table*}

We discuss influence of parameter uncertainties here.
As we can see from Eq.(\ref{Etp}), the total energy release during collisions are much larger than the total pinning energy, number of excited neutrons is also extremely larger than the total number of vortices in the crust of SGR J1935+2154.
Therefore, the actual values of $\Delta$, $E_{\rm f}$ and $k_{\rm F}$ are of little importance.
Choices of these parameters have little influence on our results.
What matters is the volume the excited neutrons occupied.
If there exists many quake sites in the equatorial plane and the strain energy is distributed isotropically around these quake sites, more vortices may be unpinned.
For our collision scenario, it is promising to induce a glitch of size larger than $10^{-6}$, depending on the surface area that asteroids/comets collide with the NS or the area that energy will be deposited when collisions happen.

\section{Conclusions and Discussions}

Inspired by the idea that FRBs from galactic magnetars may be associated with glitches, we explore the glitch size reachable for several physical processes which could result in FRBs and glitches simultaneously.
These physical processes include global starquakes in neutron stars and strangeon stars, crust fractures in magnetars, and collisions between NSs and asteroids/comets.
Our calculations are based on rotational parameters of SGR J1935+2154, as FRB 200428 from SGR J1935+2154 is the only galactic FRB detected till now.
The glitch size is constrained through the energy release during the X-ray burst and/or the SGR J1935+2154-like radio burst rate.
Summary of the glitch sizes for various physical processes and mechanisms are presented in Table \ref{glitch size summary}.

We find that, for the global starquake scenario, glitch size of spin down strained starquake for SGR J1935+2154 as a normal NS is too small to be detected, besides, the elastic energy release can not fulfill the burst energy requirement, nor will this energy induce vortex unpinning efficiently.
Glitch size of spin down strained starquake for SGR J1935+2154 as a strangeon star could be large, the energy requirement could also be simultaneously fulfilled.
However, recovery coefficient of the glitch should be low enough, for example, if a large glitch ($\Delta\nu/\nu\sim 10^{-6}$) is expected, the recovery coefficient should be as low as $Q\sim10^{-4}$.
If $Q\sim 1$, similar with those observed in other magnetars and high magnetic field pulsars, the strangeon star model will be challenged.

For the local starquake (or crust fracture) scenario, the glitch size contributed by crustal movement and moment of inertia decrease lies in the range $4.5\times 10^{-14}-4.5\times 10^{-13}$, the upper limit corresponds to the critical strain angle $\theta_{\rm cr}\simeq 0.01$.
The glitch can be amplified through vortex unpinning induced by crustal movement.
Assuming the geometry of the breaking region to be a cubic broken plate in the equatorial region, the glitch size can be amplified to $4.5\times 10^{-8}-1.8\times 10^{-7}$, the upper limit corresponds to $\theta_{\rm cr}\simeq 0.01$ again.
The glitch size may be further amplified if SGR J1935+2154 is a Vela-like mature pulsar and avalanche occurs.
From this point of view, it is important to know the exact age of SGR J1935+2154 and to assess whether it is a Crab-like or Vela-like pulsar.

For the orbital angular momentum transfer scenario, the glitch size is proportional to the mass of the asteroid/comet.
Mass of the asteroid/comet has been constrained to be $(10^{18}-6.2\times 10^{21})~\rm{g}$ by the burst energy $E_{\rm iso-X}$.
In the most optimistic case, size of the sudden glitch/anti-glitch will not be larger than $5.6\times 10^{-8}$.
If mass of the asteroid is $\sim 10^{20}~\rm{g}$ as proposed by Dai~\citep{2020ApJ...897L..40D}, the corresponding maximum glitch/anti-glitch size will be $\sim 0.9\times 10^{-9}$.
However, the collision between a NS and the asteroid/comet may deposit energy in the inner crust, resulting in vortex unpinning through heat deposition and/or neutron-vortex scattering.
For the heat deposition case, the asteroid maximum mass of $6.2\times 10^{21}~\rm{g}$ corresponds to a glitch size of $< 6.7\times 10^{-6}$ (a slow rise glitch) for a Crab-like pulsar or $<3.6\times 10^{-5}$ for a Vela-like pulsar,
while a asteroid mass of $\sim 10^{20}~\rm{g}$~~\citep{2020ApJ...897L..40D} corresponds to the  maximum glitch size of $<10^{-7}$ (a slow rise glitch for a Crab-like pulsar) or $< 10^{-6}$ (Vela-like) according to Fig.(15) in~\cite{1996ApJ...457..844L}.
For the neutron-vortex scattering case, the corresponding glitch size is $8\times 10^{-9}-1.2\times 10^{-6}$.
The upper limit corresponds to a maximum asteroid/comet mass of $6.2\times 10^{21}~\rm{g}$.
If the mass of the asteroid is $\sim 10^{20}~\rm{erg}$, the number of unpinned vortices decreases to $\sim 4.7\times 10^{-2}N_{\rm v}$ according to Eqs.(\ref{e-dep}) and (\ref{vortex number}), meanwhile, the maximum glitch size decreases to $5.6\times 10^{-8}(\Delta t_{q}/11~\rm{yr})$.
Therefore, our preliminary result shows that, the total effects of angular momentum transfer and vortex unpinning through heat deposition and neutron-vortex scattering, are unlikely to result in a large glitch with size up to $10^{-6}$ for a asteroid mass of $\sim 10^{20}~\rm{g}$.
However, the collision may affect the magnetic field configuration and induce redistribution of magnetic stresses in the crust of magnetars, in this case, the geometry of the collision site should be taken into account.
We leave this possibility to be discussed in future works.

It should be pointed out that, the glitch sizes contributed by vortex unpinning through different triggering mechanisms have no dependency on the recovery coefficient $Q$, which makes it different from the strangeon star model.

To sum up, global starquakes in strangeon stars and local starquakes (or crust fractures) in magnetars are both promising to produce large glitches with sizes up to $10^{-6}$ in association with FRB 200428, however, the former has dependency on the recovery coefficient $Q$, and $Q$ should be as low as $10^{-4}$.
While crust fractures in magnetars have no dependency on $Q$.
This difference may also serve as a criterion to distinguish strangeon star model from neutron star model if a large glitch is found to be coincident with FRBs from galactic magnetars.
The magnetar-asteroid impact model is unlikely to produce glitches with amplitudes up to $10^{-6}$ if the asteroid mass is $\sim 10^{20}~\rm{g}$.
Therefore, we conclude that, it is possible to constrain FRB models through glitch behaviors (glitch size, rise timescale, the recovery coefficient $Q$ and spin down rate offset) if glitches are accompanied by FRBs from galactic magnetars in the future.

We stress that, this work represents a very initial work, the accurate surface temperature and crustal temperature profile are needed to model the heat deposition process comprehensively.
Evolutionary age of SGR J1935+2154 should also be known to evaluate the possibility of avalanche in this pulsar.
Apart from these, the possible involvement of core superfluid in magnetars is not considered here.
The catastrophic unpinning of neutron superfluid vortex lines in the inner crust triggered by large starquake deserves careful consideration in future works~\citep{2000ASSL..254...95E}.
Besides, this work focus mainly on the FRB-glitch association, if it turns out to be an anti-glitch, much more deserves to be done.

If a glitch is detected to be in association with FRB 200428, is it reasonable to constrain FRB models with glitch behaviors?
As we konw, large radiative changes are usually accompanied by timing anomalies, but only $20-30$ percent timing anomalies are accompanied by radiative changes in magnetars~\citep{2014ApJ...784...37D,2017ARA&A..55..261K}, reflecting no direct connections between glitches/anti-glitches and outbursts.
Besides, though FRB 200428 appeared after SGR J1935+2154 went into outburst, it is unlikely that this FRB is directly correlated with the outburst or burst itself.
This can be illustrated from two aspects.
Firstly, the non-thermal X-ray burst makes it different from normal short bursts from magnetars, which prefer thermal origins~\citep{2008ApJ...685.1114I,2012ApJ...756...54L}.
Secondly, no pulsed radio emission was detected to be coincident with SGR bursts, suggesting the rare FRB-SGR burst associations~\citep{2020Natur.587...63L}.
Therefore, we believe that it is reasonable to constrain FRB models with glitch behaviors.
A much tighter constraint will be given if the glitch-FRB association is confirmed from quiescent galactic magnatars with persistent radio emission.

\section*{Acknowledgements}
We appreciate the referee for the valuable comments and suggestions, which have helped to improve this manuscript.
We are grateful to Yong Gao from KIAA for the useful discussions.
This work is supported by the National Key R\&D Program of China (Grant No. 2017YFA0402602), the National Natural Science Foundation of China (Grant Nos. 2020SKA0120300, 12003009, 12033001 and 11773011), the Strategic Priority Research Program of Chinese Academy of Sciences (Grant No. XDB23010200), the CAS ``Light of West China" Program (Grant No. 2019-XBQNXZ-B-016) and Innovation Fund of Key Laboratory of Quark and Leption Physics (Grant No. QLPL2020P01).

\section*{ORCID iDs}
Wang, W. H., https://orcid.org/0000-0003-1473-5713
\\
Xu, R. X., https://orcid.org/0000-0002-9042-3044\\
Wang, W.-Y.,
https://orcid.org/0000-0001-9036-8543

\section*{Data availability}

The data underlying this article are available in this article, no new data needed to be generated or analysed.



\begin{thebibliography}{}
\bibitem[\protect\citeauthoryear{Akbal \& Alpar}{2018}]{2018MNRAS.473..621A} Akbal O., Alpar M.~A., 2018, MNRAS, 473, 621.

\bibitem[Akbal et al. (2015)]{Akbal2015}
Akbal O., G\"ugercino\u{g}lu E., \c{S}a\c{s}maz Mu\c{s} S., Alpar M. A., 2015, Mon. Not. Roy. Astron. Soc., 449, 933

\bibitem[\protect\citeauthoryear{Alpar et al.}{1984}]{1984ApJ...276..325A} Alpar M.~A., Pines D., Anderson P.~W., Shaham J., 1984, ApJ, 276, 325.

\bibitem[\protect\citeauthoryear{Alpar, Nandkumar, \& Pines}{1985}]{1985ApJ...288..191A} Alpar M.~A., Nandkumar R., Pines D., 1985, ApJ, 288, 191.

\bibitem[\protect\citeauthoryear{Alpar et al.}{1993}]{1993ApJ...409..345A} Alpar M.~A., Chau H.~F., Cheng K.~S., Pines D., 1993, ApJ, 409, 345.

\bibitem[\protect\citeauthoryear{Alpar}{1977}]{1977ApJ...213..527A} Alpar M.~A., 1977, ApJ, 213, 527.

\bibitem[\protect\citeauthoryear{Anderson \& Itoh}{1975}]{1975Natur.256...25A} Anderson P.~W., Itoh N., 1975, Natur, 256, 25.

\bibitem[\protect\citeauthoryear{Andersson et al.}{2012}]{2012PhRvL.109x1103A} Andersson N., Glampedakis K., Ho W.~C.~G., Espinoza C.~M., 2012, PhRvL, 109, 241103.






\bibitem[\protect\citeauthoryear{Baiko \& Chugunov}{2018}]{2018MNRAS.480.5511B} Baiko D.~A., Chugunov A.~I., 2018, MNRAS, 480, 5511.

\bibitem[\protect\citeauthoryear{Bailes et al.}{2021}]{2021MNRAS.tmp..771B} Bailes M., Bassa C.~G., Bernardi G., Buchner S., Burgay M., Caleb M., Cooper A.~J., et al., 2021, MNRAS.tmp. doi:10.1093/mnras/stab749

\bibitem[\protect\citeauthoryear{Baym \& Pines}{1971}]{1971AnPhy..66..816B} Baym G., Pines D., 1971, AnPhy, 66, 816.

\bibitem[\protect\citeauthoryear{Baym, Pethick, \& Pines}{1969}]{1969Natur.224..673B} Baym G., Pethick C., Pines D., 1969, Natur, 224, 673.


\bibitem[\protect\citeauthoryear{Bochenek et al.}{2020}]{2020Natur.587...59B} Bochenek C.~D., Ravi V., Belov K.~V., Hallinan G., Kocz J., Kulkarni S.~R., McKenna D.~L., 2020, Natur, 587, 59.

\bibitem[\protect\citeauthoryear{Chamel}{2013}]{2013PhRvL.110a1101C} Chamel N., 2013, PhRvL, 110, 011101.

\bibitem[\protect\citeauthoryear{Chamel \& Haensel}{2008}]{2008LRR....11...10C} Chamel N., Haensel P., 2008, LRR, 11, 10.



\bibitem[\protect\citeauthoryear{Chawla et al.}{2020}]{2020ApJ...896L..41C} Chawla P., Andersen B.~C., Bhardwaj M., Fonseca E., Josephy A., Kaspi V.~M., Michilli D., et al., 2020, ApJL, 896, L41.

\bibitem[\protect\citeauthoryear{CHIME/FRB Collaboration et al.}{2020}]{2020Natur.582..351C} CHIME/FRB Collaboration, Amiri M., Andersen B.~C., Bandura K.~M., Bhardwaj M., Boyle P.~J., Brar C., et al., 2020, Natur, 582, 351.

\bibitem[\protect\citeauthoryear{CHIME/FRB Collaboration et al.}{2020}]{2020Natur.587...54C} CHIME/FRB Collaboration, Andersen B.~C., Bandura K.~M., Bhardwaj M., Bij A., Boyce M.~M., Boyle P.~J., et al., 2020, Natur, 587, 54.

\bibitem[\protect\citeauthoryear{Chugunov \& Horowitz}{2010}]{2010MNRAS.407L..54C} Chugunov A.~I., Horowitz C.~J., 2010, MNRAS, 407, L54.


\bibitem[\protect\citeauthoryear{Colgate \& Petschek}{1981}]{1981ApJ...248..771C} Colgate S.~A., Petschek A.~G., 1981, ApJ, 248, 771.

\bibitem[\protect\citeauthoryear{Cordes \& Shannon}{2008}]{2008ApJ...682.1152C} Cordes J.~M., Shannon R.~M., 2008, ApJ, 682, 1152.

\bibitem[\protect\citeauthoryear{Cutler, Ushomirsky, \& Link}{2003}]{2003ApJ...588..975C} Cutler C., Ushomirsky G., Link B., 2003, ApJ, 588, 975.

\bibitem[\protect\citeauthoryear{Dado \& Dar}{2020}]{2020arXiv200708370D} Dado S., Dar A., 2020, arXiv, arXiv:2007.08370

\bibitem[\protect\citeauthoryear{Dai et al.}{2016}]{2016ApJ...829...27D} Dai Z.~G., Wang J.~S., Wu X.~F., Huang Y.~F., 2016, ApJ, 829, 27.

\bibitem[\protect\citeauthoryear{Dai}{2020}]{2020ApJ...897L..40D} Dai Z.~G., 2020, ApJL, 897, L40.

\bibitem[\protect\citeauthoryear{Deibel et al.}{2015}]{2015ApJ...809L..31D} Deibel A., Cumming A., Brown E.~F., Page D., 2015, ApJL, 809, L31.

\bibitem[\protect\citeauthoryear{Dib \& Kaspi}{2014}]{2014ApJ...784...37D} Dib R., Kaspi V.~M., 2014, ApJ, 784, 37.

\bibitem[\protect\citeauthoryear{Epstein \& Link}{2000}]{2000ASSL..254...95E} Epstein R.~I., Link B., 2000, ASSL, 254, 95.

\bibitem[\protect\citeauthoryear{Espinoza et al.}{2011}]{2011MNRAS.414.1679E} Espinoza C.~M., Lyne A.~G., Stappers B.~W., Kramer M., 2011, MNRAS, 414, 1679.

\bibitem[\protect\citeauthoryear{Franco, Link, \& Epstein}{2000}]{2000ApJ...543..987F} Franco L.~M., Link B., Epstein R.~I., 2000, ApJ, 543, 987.

\bibitem[\protect\citeauthoryear{Fuentes et al.}{2017}]{2017A&A...608A.131F} Fuentes J.~R., Espinoza C.~M., Reisenegger A., Shaw B., Stappers B.~W., Lyne A.~G., 2017, A\&A, 608, A131.

\bibitem[\protect\citeauthoryear{Gao et al.}{2016}]{2016MNRAS.456...55G} Gao Z.~F., Li X.-D., Wang N., Yuan J.~P., Wang P., Peng Q.~H., Du Y.~J., 2016, MNRAS, 456, 55.

\bibitem[\protect\citeauthoryear{Gao et al.}{2019}]{2019AN....340.1030G} Gao Z.-F., Omar N., Shi X.-C., Wang N., 2019, AN, 340, 1030.

\bibitem[\protect\citeauthoryear{Geng \& Huang}{2015}]{2015ApJ...809...24G} Geng J.~J., Huang Y.~F., 2015, ApJ, 809, 24.

\bibitem[\protect\citeauthoryear{Giliberti et al.}{2019}]{2019PASA...36...36G} Giliberti E., Antonelli M., Cambiotti G., Pizzochero P.~M., 2019, PASA, 36, e036.

\bibitem[\protect\citeauthoryear{Giliberti et al.}{2020}]{2020MNRAS.491.1064G} Giliberti E., Cambiotti G., Antonelli M., Pizzochero P.~M., 2020, MNRAS, 491, 1064.

\bibitem[\protect\citeauthoryear{Gotthelf et al.}{2002}]{2002ASPC..271..309G} Gotthelf E.~V., Gavriil F.~P., Kaspi V.~M., Vasisht G., Chakrabarty D., 2002, ASPC, 271, 309

\bibitem[\protect\citeauthoryear{Fong \& Berger}{2014}]{Fong2014}
Fong W., Berger E., 2014, GRB Coordinates Network, 16542, 1

\bibitem[\protect\citeauthoryear{Galensler}{2014}]{Galensler2014}
Galensler B. M., 2014, GRB Coordinates Network, 16533, 1

\bibitem[\protect\citeauthoryear{G{\"o}{\u{g}}{\"u}{\c{s}} et al.}{2020}]{2020ApJ...905L..31G} G{\"o}{\u{g}}{\"u}{\c{s}} E., Baring M.~G., Kouveliotou C., G{\"u}ver T., Lin L., Roberts O.~J., Younes G., et al., 2020, ApJL, 905, L31.

\bibitem[\protect\citeauthoryear{G{\"u}gercino{\v{g}}lu \& Alpar}{2019}]{2019MNRAS.488.2275G} G{\"u}gercino{\v{g}}lu E., Alpar M.~A., 2019, MNRAS, 488, 2275.

\bibitem[\protect\citeauthoryear{Haskell et al.}{2008}]{2008MNRAS.385..531H} Haskell B., Samuelsson L., Glampedakis K., Andersson N., 2008, MNRAS, 385, 531.

\bibitem[\protect\citeauthoryear{Ho et al.}{2015}]{2015SciA....1E0578H} Ho W.~C.~G., Espinoza C.~M., Antonopoulou D., Andersson N., 2015, SciA, 1, e1500578.

\bibitem[\protect\citeauthoryear{Hoffman \& Heyl}{2012}]{2012MNRAS.426.2404H} Hoffman K., Heyl J., 2012, MNRAS, 426, 2404.

\bibitem[\protect\citeauthoryear{Horowitz \& Kadau}{2009}]{2009PhRvL.102s1102H} Horowitz C.~J., Kadau K., 2009, PhRvL, 102, 191102.

\bibitem[\protect\citeauthoryear{Huang \& Geng}{2014}]{2014ApJ...782L..20H} Huang Y.~F., Geng J.~J., 2014, ApJL, 782, L20.


\bibitem[\protect\citeauthoryear{Israel et al.}{2016}]{2016MNRAS.457.3448I} Israel G.~L., Esposito P., Rea N., Coti Zelati F., Tiengo A., Campana S., Mereghetti S., et al., 2016, MNRAS, 457, 3448.

\bibitem[\protect\citeauthoryear{Israel et al.}{2008}]{2008ApJ...685.1114I} Israel G.~L., Romano P., Mangano V., Dall'Osso S., Chincarini G., Stella L., Campana S., et al., 2008, ApJ, 685, 1114.

\bibitem[\protect\citeauthoryear{Jones \& Andersson}{2001}]{2001MNRAS.324..811J} Jones D.~I., Andersson N., 2001, MNRAS, 324, 811.

\bibitem[\protect\citeauthoryear{Kaminker et al.}{2009}]{2009MNRAS.395.2257K} Kaminker A.~D., Potekhin A.~Y., Yakovlev D.~G., Chabrier G., 2009, MNRAS, 395, 2257.

\bibitem[\protect\citeauthoryear{Kaspi \& Beloborodov}{2017}]{2017ARA&A..55..261K} Kaspi V.~M., Beloborodov A.~M., 2017, ARA\&A, 55, 261.

\bibitem[\protect\citeauthoryear{Katz, Toole, \& Unruh}{1994}]{1994ApJ...437..727K} Katz J.~I., Toole H.~A., Unruh S.~H., 1994, ApJ, 437, 727.

\bibitem[\protect\citeauthoryear{Katz}{2018}]{2018PrPNP.103....1K} Katz J.~I., 2018, PrPNP, 103, 1.

\bibitem[\protect\citeauthoryear{Kothes et al.}{2018}]{2018ApJ...852...54K} Kothes R., Sun X., Gaensler B., Reich W., 2018, ApJ, 852, 54.

\bibitem[\protect\citeauthoryear{Kumar et al.}{2014}]{2014ApJ...781...41K} Kumar H.~S., Safi-Harb S., Slane P.~O., Gotthelf E.~V., 2014, ApJ, 781, 41.

\bibitem[\protect\citeauthoryear{Layek \& Yadav}{2020}]{2020MNRAS.499..455L} Layek B., Yadav P.~R., 2020, MNRAS, 499, 455.


\bibitem[\protect\citeauthoryear{Lai et al.}{2018}]{2018MNRAS.476.3303L} Lai X.~Y., Yun C.~A., Lu J.~G., L{\"u} G.~L., Wang Z.~J., Xu R.~X., 2018, MNRAS, 476, 3303.

\bibitem[\protect\citeauthoryear{Lander et al.}{2015}]{2015MNRAS.449.2047L} Lander S.~K., Andersson N., Antonopoulou D., Watts A.~L., 2015, MNRAS, 449, 2047.

\bibitem[\protect\citeauthoryear{Li et al.}{2016}]{2016ApJS..223...16L} Li A., Dong J.~M., Wang J.~B., Xu R.~X., 2016, ApJS, 223, 16.

\bibitem[\protect\citeauthoryear{Li et al.}{2021}]{2021NatAs.tmp...48L} Li C.~K., Lin L., Xiong S.~L., Ge M.~Y., Li X.~B., Li T.~P., Lu F.~J., et al., 2021, NatAs.tmp. doi:10.1038/s41550-021-01302-6

\bibitem[\protect\citeauthoryear{Lin et al.}{2012}]{2012ApJ...756...54L} Lin L., G{\"o}{\v{g}}{\"u}{\c{s}} E., Baring M.~G., Granot J., Kouveliotou C., Kaneko Y., van der Horst A., et al., 2012, ApJ, 756, 54.

\bibitem[\protect\citeauthoryear{Lin et al.}{2020}]{2020Natur.587...63L} Lin L., Zhang C.~F., Wang P., Gao H., Guan X., Han J.~L., Jiang J.~C., et al., 2020, Natur, 587, 63.

\bibitem[\protect\citeauthoryear{Link, Epstein, \& Lattimer}{1999}]{1999PhRvL..83.3362L} Link B., Epstein R.~I., Lattimer J.~M., 1999, PhRvL, 83, 3362.

\bibitem[\protect\citeauthoryear{Link \& Epstein}{1996}]{1996ApJ...457..844L} Link B., Epstein R.~I., 1996, ApJ, 457, 844.

\bibitem[\protect\citeauthoryear{Lorimer et al.}{2007}]{2007Sci...318..777L} Lorimer D.~R., Bailes M., McLaughlin M.~A., Narkevic D.~J., Crawford F., 2007, Sci, 318, 777.

\bibitem[\protect\citeauthoryear{Mereghetti et al.}{2020}]{2020ApJ...898L..29M} Mereghetti S., Savchenko V., Ferrigno C., G{\"o}tz D., Rigoselli M., Tiengo A., Bazzano A., et al., 2020, ApJL, 898, L29.

\bibitem[\protect\citeauthoryear{Mitrofanov \& Sagdeev}{1990}]{1990Natur.344..313M} Mitrofanov I.~G., Sagdeev R.~Z., 1990, Natur, 344, 313.

\bibitem[\protect\citeauthoryear{Pastore, Baroni, \& Losa}{2011}]{2011PhRvC..84f5807P} Pastore A., Baroni S., Losa C., 2011, PhRvC, 84, 065807.

\bibitem[\protect\citeauthoryear{Peng \& Xu}{2008}]{2008MNRAS.384.1034P} Peng C., Xu R.~X., 2008, MNRAS, 384, 1034.

\bibitem[\protect\citeauthoryear{Petroff et al.}{2019}]{2019A&ARv..27....4P} Petroff E., Hessels J.~W.~T., Lorimer D.~R., 2019, A\&ARv, 27, 4.

\bibitem[\protect\citeauthoryear{Pons, Miralles, \& Geppert}{2009}]{2009A&A...496..207P} Pons J.~A., Miralles J.~A., Geppert U., 2009, A\&A, 496, 207.

\bibitem[\protect\citeauthoryear{Rencoret, Aguilera-G{\'o}mez, \& Reisenegger}{2021}]{2021arXiv210612604R} Rencoret J.~A., Aguilera-G{\'o}mez C., Reisenegger A., 2021, arXiv, arXiv:2106.12604

\bibitem[\protect\citeauthoryear{Ridnaia et al.}{2021}]{2021NatAs.tmp...30R} Ridnaia A., Svinkin D., Frederiks D., Bykov A., Popov S., Aptekar R., Golenetskii S., et al., 2021, NatAs.tmp. doi:10.1038/s41550-020-01265-0

\bibitem[\protect\citeauthoryear{Ruderman}{1969}]{1969Natur.223..597R} Ruderman M., 1969, Natur, 223, 597.

\bibitem[\protect\citeauthoryear{Ruderman}{1991}]{1991ApJ...382..587R} Ruderman M., 1991, ApJ, 382, 587.

\bibitem[\protect\citeauthoryear{Ruderman, Zhu, \& Chen}{1998}]{1998ApJ...492..267R} Ruderman M., Zhu T., Chen K., 1998, ApJ, 492, 267.

\bibitem[\protect\citeauthoryear{Sinha \& Sedrakian}{2015}]{2015PhRvC..91c5805S} Sinha M., Sedrakian A., 2015, PhRvC, 91, 035805.

\bibitem[\protect\citeauthoryear{Spitler et al.}{2016}]{2016Natur.531..202S} Spitler L.~G., Scholz P., Hessels J.~W.~T., Bogdanov S., Brazier A., Camilo F., Chatterjee S., et al., 2016, Natur, 531, 202.

\bibitem[\protect\citeauthoryear{Suvorov \& Kokkotas}{2019}]{2019MNRAS.488.5887S} Suvorov A.~G., Kokkotas K.~D., 2019, MNRAS, 488, 5887.

\bibitem[\protect\citeauthoryear{Tavani et al.}{2021}]{2021NatAs.tmp...31T} Tavani M., Casentini C., Ursi A., Verrecchia F., Addis A., Antonelli L.~A., Argan A., et al., 2021, NatAs.tmp. doi:10.1038/s41550-020-01276-x

\bibitem[\protect\citeauthoryear{Thompson, Lyutikov, \& Kulkarni}{2002}]{2002ApJ...574..332T} Thompson C., Lyutikov M., Kulkarni S.~R., 2002, ApJ, 574, 332.

\bibitem[\protect\citeauthoryear{Thompson \& Duncan}{1995}]{1995MNRAS.275..255T} Thompson C., Duncan R.~C., 1995, MNRAS, 275, 255.

\bibitem[\protect\citeauthoryear{Thompson \& Duncan}{1996}]{1996ApJ...473..322T} Thompson C., Duncan R.~C., 1996, ApJ, 473, 322.

\bibitem[\protect\citeauthoryear{Thornton et al.}{2013}]{2013Sci...341...53T} Thornton D., Stappers B., Bailes M., Barsdell B., Bates S., Bhat N.~D.~R., Burgay M., et al., 2013, Sci, 341, 53.

\bibitem[\protect\citeauthoryear{Wang et al.}{2020}]{2020MNRAS.499..355W} Wang W.-Y., Zhang B., Chen X., Xu R., 2020, MNRAS, 499, 355.


\bibitem[\protect\citeauthoryear{Vigan{\`o} et al.}{2013}]{2013MNRAS.434..123V} Vigan{\`o} D., Rea N., Pons J.~A., Perna R., Aguilera D.~N., Miralles J.~A., 2013, MNRAS, 434, 123.



\bibitem[\protect\citeauthoryear{Wang et al.}{2021}]{2021MNRAS.500.5336W} Wang W.~H., Lai X.~Y., Zhou E.~P., Lu J.~G., Zheng X.~P., Xu R.~X., 2021, MNRAS, 500, 5336.

\bibitem[\protect\citeauthoryear{Wang et al.}{2018}]{2018ApJ...852..140W} Wang W., Luo R., Yue H., Chen X., Lee K., Xu R., 2018, ApJ, 852, 140.

\bibitem[\protect\citeauthoryear{Watanabe \& Pethick}{2017}]{2017PhRvL.119f2701W} Watanabe G., Pethick C.~J., 2017, PhRvL, 119, 062701.

\bibitem[\protect\citeauthoryear{Wlaz{\l}owski et al.}{2016}]{2016PhRvL.117w2701W} Wlaz{\l}owski G., Sekizawa K., Magierski P., Bulgac A., Forbes M.~M., 2016, PhRvL, 117, 232701.

\bibitem[\protect\citeauthoryear{Xu}{2003}]{2003ApJ...596L..59X} Xu R.~X., 2003, ApJL, 596, L59.

\bibitem[\protect\citeauthoryear{Yang \& Zhang}{2021}]{2021arXiv210401925Y} Yang Y.-P., Zhang B., 2021, arXiv, arXiv:2104.01925

\bibitem[\protect\citeauthoryear{Younes et al.}{2020}]{2020ApJ...904L..21Y} Younes G., G{\"u}ver T., Kouveliotou C., Baring M.~G., Hu C.-P., Wadiasingh Z., Begi{\c{c}}arslan B., et al., 2020, ApJL, 904, L21.

\bibitem[\protect\citeauthoryear{Yuan et al.}{2010}]{2010ApJ...719L.111Y} Yuan J.~P., Manchester R.~N., Wang N., Zhou X., Liu Z.~Y., Gao Z.~F., 2010, ApJL, 719, L111.

\bibitem[\protect\citeauthoryear{Zdunik, Bejger, \& Haensel}{2008}]{2008A&A...491..489Z} Zdunik J.~L., Bejger M., Haensel P., 2008, A\&A, 491, 489.

\bibitem[\protect\citeauthoryear{Zhang}{2020}]{2020Natur.587...45Z} Zhang B., 2020, Natur, 587, 45.
\bibitem[\protect\citeauthoryear{Zhou et al.}{2004}]{2004APh....22...73Z} Zhou A.~Z., Xu R.~X., Wu X.~J., Wang N., 2004, APh, 22, 73.

\bibitem[\protect\citeauthoryear{Zhou et al.}{2014}]{2014MNRAS.443.2705Z} Zhou E.~P., Lu J.~G., Tong H., Xu R.~X., 2014, MNRAS, 443, 2705.


\bibitem[\protect\citeauthoryear{Zhou et al.}{2020}]{2020ApJ...905...99Z} Zhou P., Zhou X., Chen Y., Wang J.-S., Vink J., Wang Y., 2020, ApJ, 905, 99.






\end{thebibliography}
\end{document}